\documentclass[twocolumn]{aastex62}

\newcommand{\HI}{\mathrm{H\,I}}
\newcommand{\HII}{\mathrm{H\,II}}
\newcommand{\HeI}{\mathrm{He\,I}}
\newcommand{\HeII}{\mathrm{He\,II}}
\newcommand{\HeIII}{\mathrm{He\,III}}

\newcommand{\lya}{Ly$\alpha$ }

\begin{document}
\title{\bf \large Quantitative Constraints on the Reionization History from the IGM Damping Wing Signature in Two Quasars at $z > 7$}

\author{Frederick B. Davies}
\affiliation{Department of Physics, University of California, Santa Barbara, CA 93106-9530, USA}

\author{Joseph F. Hennawi}
\affiliation{Department of Physics, University of California, Santa Barbara, CA 93106-9530, USA}

\author{Eduardo Ba\~{n}ados}
\altaffiliation{Carnegie-Princeton Fellow}
\affiliation{The Observatories of the Carnegie Institution for Science, 813 Santa Barbara Street, Pasadena, California 91101, USA}

\author{Zarija Luki\'{c}}
\affiliation{Lawrence Berkeley National Laboratory, CA 94720-8139, USA}

\author{Roberto Decarli}
\affiliation{Max Planck Institut f\"{u}r Astronomie, K\"{o}nigstuhl 17, D-69117 Heidelberg, Germany}
\affiliation{INAF--Osservatorio Astronomico di Bologna, via Gobetti 93/3, 40129 Bologna, Italy}

\author{Xiaohui Fan}
\affiliation{Steward Observatory, The University of Arizona, 933 North Cherry Avenue, Tucson, Arizona 85721-0065, USA}

\author{Emanuele P. Farina}
\affiliation{Department of Physics, University of California, Santa Barbara, CA 93106-9530, USA}

\author{Chiara Mazzucchelli}
\affiliation{Max Planck Institut f\"{u}r Astronomie, K\"{o}nigstuhl 17, D-69117 Heidelberg, Germany}

\author{Hans-Walter Rix}
\affiliation{Max Planck Institut f\"{u}r Astronomie, K\"{o}nigstuhl 17, D-69117 Heidelberg, Germany}

\author{Bram P. Venemans}
\affiliation{Max Planck Institut f\"{u}r Astronomie, K\"{o}nigstuhl 17, D-69117 Heidelberg, Germany}

\author{Fabian Walter}
\affiliation{Max Planck Institut f\"{u}r Astronomie, K\"{o}nigstuhl 17, D-69117 Heidelberg, Germany}

\author{Feige Wang}
\affiliation{Department of Physics, University of California, Santa Barbara, CA 93106-9530, USA}
\affiliation{Department of Astronomy, School of Physics, Peking University, Beijing 100871, China}
\affiliation{Kavli Institute for Astronomy and Astrophysics, Peking University, Beijing 100871, China}

\author{Jinyi Yang}
\affiliation{Steward Observatory, The University of Arizona, 933 North Cherry Avenue, Tucson, Arizona 85721-0065, USA}
\affiliation{Department of Astronomy, School of Physics, Peking University, Beijing 100871, China}
\affiliation{Kavli Institute for Astronomy and Astrophysics, Peking University, Beijing 100871, China}

\begin{abstract}
  During reionization, neutral hydrogen in the intergalactic medium
  (IGM) imprints a damping wing absorption feature on the spectrum of
  high-redshift quasars. A detection of this signature provides
  compelling evidence for a significantly neutral Universe, and
  enables measurements of the hydrogen neutral fraction $x_{\rm
    HI}(z)$ at that epoch.  Obtaining reliable quantitative
  constraints from this technique, however, is challenging due
  stochasticity induced by the patchy inside-out topology of
  reionization, degeneracies with quasar lifetime, and the unknown
  unabsorbed quasar spectrum close to rest-frame Ly$\alpha$. We
  combine a large-volume semi-numerical simulation of reionization
  topology with 1D radiative transfer through high-resolution
  hydrodynamical simulations of the high-redshift Universe to
  construct models of quasar transmission spectra during
  reionization. Our state-of-the-art approach captures the
  distribution of damping wing strengths in biased quasar halos that
  should have reionized earlier, as well as the erosion of neutral gas
  in the quasar environment caused by its own ionizing radiation. Combining this
  detailed model with our new technique for predicting the quasar continuum and
  its associated uncertainty, we introduce a Bayesian statistical
  method to jointly constrain the neutral fraction of the Universe and
  the quasar lifetime from individual quasar spectra. We apply this methodology
  to the spectra of the two highest redshift quasars known, ULAS
  J1120+0641 and ULAS J1342+0928, and measured volume-averaged neutral
  fractions $\langle x_{\rm HI} \rangle(z=7.09)=0.48^{+0.26}_{-0.26}$
  and $\langle x_{\rm HI} \rangle(z=7.54)=0.60^{+0.20}_{-0.23}$
  (posterior medians and 68\% credible intervals) when marginalized
  over quasar lifetimes of $10^3 \leq t_{\rm q} \leq 10^8$ years.
\end{abstract}

\section{Introduction}

The epoch of reionization was a landmark event in the history of the Universe when the cumulative number of ionizing photons escaping from the first stars, galaxies, and quasars surpassed the number of hydrogen atoms in the intergalactic medium (IGM). Our knowledge of reionization is bounded by the presence of transmission in the Ly$\alpha$ forest at $z\la6$ \citep{Fan06}, and an integral constraint from the electron scattering optical depth of the cosmic microwave background (CMB) which constrains the volume of ionized IGM between the present day and $z\sim1100$ \citep{Planck16a} that suggests a characteristic reionization redshift of $z_{\rm re}=6.4$--$9.7$ (95\% credible interval, \citealt{Planck16b}). With only these constraints, the detailed reionization history -- reflecting the nature and evolution of sources of ionizing photons -- is still highly uncertain and model-dependent.

The discovery and deep follow-up spectroscopy of quasars with redshifts greater than six
provided the first look at the IGM approaching the epoch of reionization (e.g. \citealt{Fan01,Fan03,Becker01,White03}). While Gunn-Peterson troughs \citep{GP65} in the Ly$\alpha$ and Ly$\beta$ forests of these quasars due to the presence of neutral hydrogen in the IGM may be signatures of ongoing reionization, they can only place lower limits on the volume-averaged hydrogen neutral fraction of
$\langle x_\HI\rangle \ga 10^{-4}$ (e.g. \citealt{Fan06}).
The sizes of the transparent proximity zones of these quasars have also been analyzed in the context of expanding Str{\"{o}}mgren spheres in a \mbox{(partially-)neutral} IGM \citep{CH00,Wyithe05,MH07,Schroeder13}, but as recently demonstrated by \citet{Eilers17}, the sizes alone may be
insensitive to the ionization state of the IGM. A more sensitive, and perhaps definitive, probe of neutral gas in the IGM is the \lya damping wing \citep{ME98} which suppresses the quasar continuum redward of rest-frame Ly$\alpha$.

 The first 
 quasar with a claimed damping wing signal was ULAS J1120+0641 \citep{Mortlock11} at $z=7.09$, but the inferred constraints on $\langle x_\HI\rangle$ vary between different analyses. These differences are in part due to differences in physical models for the proximity zone and/or damping wing.
 One approach to constrain $\langle x_{\rm HI}\rangle$ is to fit the Ly$\alpha$ transmission spectrum with the analytic model of \citet{ME98}, as performed by \citet{Mortlock11}, but this formula does not include the substantial resonant Ly$\alpha$ absorption by residual \ion{H}{1} inside of the proximity zone (see, e.g., \citealt{Bolton11,Keating15}), and the IGM outside the proximity zone is typically assumed to have a completely uniform ionization state instead of a more realistic patchy topology of ionized bubbles \citep{Furlanetto04}. \citet{Greig17b} constrained $\langle x_{\rm HI}\rangle$ from the damping wing of ULAS J1120+0641 using large-volume semi-numerical simulations of the reionization topology \citep{Mesinger16} to predict the distribution of damping wing strengths as a function of $\langle x_{\rm HI}\rangle$, but they only considered wavelengths redward of Ly$\alpha$. The size of the proximity zone, and the strength of the damping wing, are sensitive to the quasar lifetime \citep{Bolton11,Keating15} which has an uncertainty of several orders of magnitude \citep{Martini04,Eilers17}.

Another complication is that uncertainties and differences between methodologies for estimating the intrinsic quasar continuum \citep{KH09,Greig17a} can be similar in strength to the damping wing signal itself. Further exacerbating this challenge is the fact that the spectral properties of quasars at $z\ga6.5$, in particular their \ion{C}{4} emission line blueshifts, are often extreme outliers of the distribution of lower redshift quasars \citep{Mazzucchelli17}. In light of its large \ion{C}{4} blueshift, \citet{Mortlock11} estimated the intrinsic spectrum of ULAS J1120+0641 via stacking a sample of SDSS quasar spectra with matched \ion{C}{4} emission line properties, but did not test the accuracy of this continuum estimation method on quasars with known continua. \citet{BB15} found that a more closely matched sample of lower-redshift quasars (i.e., without any damping wing signal) with carefully selected \ion{C}{4} emission line strengths and blueshifts had Ly$\alpha$ spectral shapes consistent with the observed ULAS J1120+0641 spectrum. In contrast, the predictive continuum model of \citet{Greig17a}, as demonstrated by \citet{Greig17b}, appears to prefer a much stronger intrinsic Ly$\alpha$ profile, suggesting instead that the damping wing signal is quite strong. 

Analysis of the recently discovered quasar ULAS J1342+0928 \citep{Banados18} at $z=7.54$ suggests that it exhibits a much more prominent damping wing absorption signal than ULAS J1120+0641, consistent with a predominantly neutral IGM. However, its \ion{C}{4} line exhibits a blueshift more than twice that of ULAS J1120+0641, and so only a very small number of similar quasars exist in lower redshift samples. \citet{Banados18} estimated the intrinsic spectrum of ULAS J1342+0928 by constructing a composite spectrum from 46 SDSS/BOSS quasars with similar \ion{C}{4} emission line properties and estimated the uncertainty via measuring the residuals between the composite and its constituent quasar spectra. They then derived their fiducial constraints on $\langle x_{\rm HI}\rangle$ in the surrounding IGM using the \citet{ME98} model for the damping wing shape. In this work we describe one of the alternative models (``Model B") from \citet{Banados18} in more detail.

A complete model of the proximity zone and damping wing region of quasar spectra requires an estimate of the intrinsic quasar continuum, the uncertainty in the quasar continuum model, a model for the small-scale density fluctuations in the IGM, a realistic description of patchy reionization topology surrounding the massive dark matter halos that host luminous quasars, and time-dependent radiative transfer of ionizing photons from the quasar along the line of sight. The goal of this work is to put all of these pieces together for the first time to forward model mock quasar spectra and develop a statistical method to constrain the volume-averaged IGM neutral fraction and quasar lifetime from individual quasar spectra. 

In \citet[][henceforth Paper I]{Davies18a}, we developed a Principal Component Analysis (PCA)-based approach with a training set of $>10,000$ quasar spectra from the SDSS/BOSS DR12Q catalog \citep{Paris17} to predict the ``blue-side" quasar continuum, at rest-frame wavelengths $1175 < \lambda_{\rm rest}<1280$ {\AA}, from the ``red-side" spectrum, covering $1280 < \lambda_{\rm rest} < 2850$ {\AA}.  We quantified the covariant uncertainties by testing the method on the training set, finding that for a typical quasar the relative error of our predicted continua is $\sim6$--$12\%$ at
rest-frame wavelengths most sensitive to damping wing absorption. Finally, we demonstrated the applicability of our method on the two known $z>7$ quasars: ULAS J1120+0641, and ULAS J1342+0928. While these quasars represent outliers from the distribution of typical quasars in SDSS/BOSS,
we have calibrated the uncertainty on the blue-side predicted continua from custom subsets of ``nearest-neighbor" quasars in the training set that have similar red-side spectra to each quasar separately.

In this work, we present a hybrid model for quasar proximity zone and damping wing structures during reionization, and a statistical method to perform Bayesian parameter inference on high-redshift quasar spectra in conjunction with the PCA quasar continuum model from Paper I. We apply these new methods to constrain $\langle x_{\rm HI}\rangle$ from the spectra of ULAS J1342+0928 at $z=7.54$ and ULAS J1120+0641 at $z=7.09$. We find strong evidence for a substantially neutral IGM at $z>7$, especially at $z=7.54$, consistent with the latest constraints from the CMB \citep{Planck16b}.

The rest of the paper is structured as follows. In \S~2, we briefly summarize the Principal Component Analysis (PCA) method for predicting the intrinsic blue-side quasar continuum from the red-side spectrum from Paper I. In \S~3 we describe our hybrid model for quasar proximity zones and the Ly$\alpha$ damping wing, combining ionizing radiative transfer simulations \citep{Davies16} through density field skewers from high-resolution hydrodynamical simulations with semi-numerical simulations of the inside-out reionization topology around massive halos. In \S~4 we describe our methodology for performing Bayesian parameter inference 
from millions of forward-modeled mock spectra. In \S~5 we show the results of our analysis on the two quasars known at $z>7$: ULAS J1342+0928 and ULAS J1120+0641. Finally, in \S~6 we conclude with a discussion of the implications of the neutral fraction constraints from the two
quasars on the reionization history of the Universe, and describe avenues for future investigation of existing quasar samples.

In this work we assume a flat $\Lambda$CDM cosmology with $h=0.685$, $\Omega_b=0.047$, $\Omega_m=0.3$, $\Omega_\Lambda=0.7$, and $\sigma_8=0.8$.

\section{PCA Continuum Model}\label{sec:pca}

We adopt the method for predicting the intrinsic blue-side quasar continuum ($1175 < \lambda_{\rm rest} < 1280$ {\AA}) from the observed red-side spectrum ($1280 < \lambda_{\rm rest} < 2850$ {\AA}) from Paper I, which we briefly summarize below.

To construct the PCA model, we selected a sample of $12,764$ quasars from the BOSS DR12Q catalog \citep{Paris17} at $2.09 < z_{\rm pipe} < 2.51$ with ${\rm S/N}>7$
at $\lambda_{\rm rest}=1290$ {\AA}, and fit each spectrum with an automated, piecewise spline fitting method designed to recover smooth quasar continua in the presence of absorption lines \citep{Young79,Carswell82,Dall'Aglio08}. In this redshift range, the BOSS spectra cover the entire spectral range from Ly$\alpha$ to \ion{Mg}{2}.
We further processed the splined spectra by median stacking each one with its 40 nearest neighbors to clean up residual artifacts such as strong associated absorption.  We then computed principal component spectra (or ``basis spectra") from these median stacks
of these spline fit spectra with the standard PCA approach using \texttt{scikit-learn} \citep{scikit-learn}, albeit in \emph{log-space}. That is, the logarithm of each quasar spectrum is represented by a sum of basis spectra ${\bf A}_i$ with corresponding weights $a_i$,
\begin{equation}
\log{\bf F} \approx \langle\log{\bf F}\rangle + \sum_i^{n} a_i {\bf A}_i,
\end{equation}
which in linear space becomes a \emph{product} of basis spectra raised to powers,
\begin{equation}\label{eqn:pca}
{\bf F} \approx {\rm e}^{\langle\log{\bf F}\rangle} \prod_i^{n} {\rm e}^{a_i {\bf A}_i}.
\end{equation}
This log-space decomposition naturally accounts for the continuum slope variations between quasars, which dominates the total variance in flux space. For our analysis, we decomposed the red-side and blue-side spectra independently, keeping 10 red-side (${\bf R}_i$) and 6 blue-side (${\bf B}_i$) basis spectra.

We then found the best-fit red-side coefficients $r_i$ for each (original, not spline fit)
quasar spectrum in the training set via $\chi^2$ minimization while simultaneously fitting for a \emph{template redshift} $z_{\rm temp}$, allowing us to place each quasar onto a consistently defined rest-frame. The blue-side coefficients $b_i$ for each training set quasar were then found by fitting the blue-side (in the $z_{\rm temp}$ frame) spline fit continua assuming constant noise. From the sets of $r_i$ and $b_i$ for all training set quasars, ${\bf r}$ and ${\bf b}$, we follow \citet{Suzuki05} and \citet{Paris11} and compute the \emph{projection matrix} ${\bf X}$ by finding the least-squares solution to the linear equation,
\begin{equation}
{\bf b} = {\bf r} \cdot {\bf X}.
\end{equation}
After fitting the 10 $r_i$ of an arbitrary quasar spectrum, we can ``project" to the corresponding 6 $b_i$ (and thus reconstruct the blue-side spectrum) via a dot product with the $10 \times 6$ projection matrix ${\bf X}$.

By testing our PCA procedure on the training set, we found that the relative error in the projected blue-side continua (which we refer to as the ``blue-side prediction") is
$\sim6$--$12\%$ in the region of the spectrum most useful for proximity zone and damping wing analyses ($1210 < \lambda_{\rm rest} < 1240$), with a mean bias $\la1\%$. The continuum error was found to be highly covariant across wide regions of the spectrum corresponding to regions associated with broad emission lines. However, as mentioned above, the spectra of $z\ga6.5$ quasars are known to be irregular compared to typical quasar spectra at lower redshift -- in particular, they exhibit large \ion{C}{4} blueshifts relative to lower ionization lines in the spectrum such as \ion{Mg}{2} \citep{Mortlock11,Mazzucchelli17,Banados18}. The uncertainty of the predicted continua for these atypical spectra may not be properly represented by the average uncertainty for all spectra.

For individual quasars, we can estimate a more accurate uncertainty by measuring the continuum errors for quasars with similar red-side spectra. We defined a distance $D_r$ in the space of red-side PCA coefficients $r_i$ by 
\begin{equation}\label{eqn:dist}
D_r \equiv \sqrt{\sum_i^{N_{\rm PCA,r}}\left(\frac{{\Delta}r_i}{\sigma(r_i)}\right)^2},
\end{equation}
where $N_{\rm PCA,r}$ is the number of red-side PCA basis vectors, ${\Delta}r_i$ is the difference between $r_i$ values, and $\sigma(r_i)$ is the standard deviation of $r_i$ values in the training set. In Paper I, we measured the predicted continuum errors for the $1\%$ of training set spectra with the lowest $D_r$ to each of the $z>7$ quasars, allowing us to estimate a custom continuum uncertainty for each $z>7$ quasar.
The predictions for these ``similar" quasars tended to be slightly less uncertain and somewhat more biased than the training set as a whole.

For the statistical analysis of quasar spectra that follows, we require the ability to generate mock realizations of the continuum prediction error, which we denote as $\epsilon_C$ following Paper I. We assume a multivariate Gaussian distribution for the relative continuum error, with the mean and covariance determined from the prediction errors measured for similar, i.e. the $1\%$ nearest neighbor,
quasars as described above. We then use draws from these custom error distributions to generate forward-modeled mock spectra, described in more detail in \S~\ref{sec:stats}.

\section{Hybrid Model of Quasar Proximity Zones and Damping Wings During Reionization}

With our predictive model for intrinsic quasar continua and their errors
in place, we now must develop a physical model for quasar proximity zones and damping wings. We construct a hybrid model with three parts:

\begin{enumerate}
\item High-resolution density field from a large-volume hydrodynamical simulation \citep{Lukic15}.
\item Semi-numerical simulations of reionization morphology (\citealt{Mesinger11}; Davies \& Furlanetto, in prep.).
\item One-dimensional ionizing radiative transfer of hydrogen- and helium-ionizing photons emitted by the quasar \citep{Davies16}.
\end{enumerate}

In this section, we describe these model components in detail.

\subsection{Nyx Hydrodynamical Simulation}

The first ingredient of our hybrid model is the small-scale structure of the IGM, which determines the absorption features inside the quasar proximity zone. We use density, velocity, and temperature fields from the $z=7.0$ output of a \texttt{Nyx} hydrodynamical simulation \citep{Almgren13}, 100 Mpc$/h$ (comoving) on a side with $4096^3$ dark matter particles and $4096^3$ baryon grid cells (see also \citealt{Lukic15}). Dark matter halos were selected via an algorithm that finds topologically-connected regions above 138 times the mean density (Luki\`{c} et al., in prep.), which is described in \citet{Sorini17}.
We extract 1200 axis-aligned skewers from the centers of the 200 most massive halos, corresponding to halo masses $M_{\rm h}\ga2\times10^{11}$ M$_\odot$. 
The simulation, optimized for studying the Ly$\alpha$ forest, was run on a fixed, Eulerian grid, and lacks prescriptions for star formation or feedback \citep{Lukic15} which are required to characterize the circumgalactic medium of massive dark matter halos.
Nevertheless, they should be adequate for our purposes, because our primary goal is to capture the larger-scale overdensity surrounding these halos on the relatively large $\sim1$--$2$ proper Mpc scales  covered by the proximity zones of the $z>7$ quasars in our analysis (compared to the halo virial radius, $\sim50$ proper kpc).

We re-scale the gas density of the skewers by $(1+z)^3$ depending on which quasar we are simulating. We leave the computation of custom-redshift outputs (i.e. matched to the quasar redshifts) of large hydrodynamical simulations to future work, but note that between $z=7$ and the redshifts of the two quasars we focus on here ($z=7.09,7.54$) the evolution of the overdensity field should be relatively unimportant.

\subsection{Semi-Numerical Reionization Simulations with \texttt{21cmFAST}}

The second ingredient of our hybrid model is the large-scale morphology of reionization around massive quasar-hosting halos. To compute realistic ionization fields on large scales, we adopt a modified version of the semi-numerical reionization code \texttt{21cmFAST}\footnote{\url{https://github.com/andreimesinger/21cmFAST}} \citep{Mesinger11}, to be presented in further detail in Davies \& Furlanetto (in prep.).
The \texttt{21cmFAST} code computes the fraction of material that has collapsed into dark matter halos, $f_{\rm coll}$, following conditional Press-Schechter \citep{LC93} applied to a non-linear density
field computed using the Zel'dovich approximation \citep{Zel'dovich70}. A region is considered ionized if $f_{\rm coll} > \zeta^{-1}$ on \emph{any} scale, where $\zeta$ is the ``ionizing efficiency," combining a series of assumptions about the efficiency of star formation and the production (and escape) of ionizing photons from galaxies into a single parameter that corresponds to the total number of ionizing photons emitted per collapsed baryon. In standard \texttt{21cmFAST}, this criterion is assessed by filtering the density field from large to small scales, re-computing $f_{\rm coll}$ at each filter scale. Our modified algorithm assigns collapsed mass to each cell according to the non-linear density field, and it is this collapsed mass field which is filtered to determine whether a given region is ionized, similar to \texttt{DexM} \citep{MF07} but without explicitly generating a distribution of halos. In this way, the small-scale clustering of halos is better reflected on large scales, and the new algorithm produces ionization fields that are very similar to \texttt{DexM} at a very small fraction of the computation time. We have also implemented a novel approach to treat the mean free path of ionizing photons as a smooth attenuation rather than a sharp cutoff.
As shown by \citet{Greig17b}, we do not expect the exact choice of model for reionization topology to make a substantial difference in our inference of $\langle x_{\rm HI} \rangle$, so we leave an exploration of different model assumptions to future work.

The hydrodynamical simulation is likely to be too small to fully characterize the distribution of ionized regions around rare, massive halos, so we compute the ionization fields in an independent larger volume, 400 comoving Mpc (cMpc) on a side. The resolution of the cosmological initial conditions was $2048^3$, while the evolved density field and ionization fields were output at a lower resolution, $512^3$. We assume a mass-independent ionizing efficiency $\zeta$, a minimum halo mass of $M_{\rm min}=10^8$ M$_\odot$, and mean free path of ionizing photons $\lambda_{\rm mfp}=60$ cMpc. We tuned $\zeta$ to produce ionization fields with global volume-averaged neutral fractions of $\langle x_{\rm HI} \rangle=0.05$--$0.95$ in steps of ${\Delta}x_{\rm HI}=0.05$.

\begin{figure}[htb]
\begin{center}
\resizebox{8.50cm}{!}{\includegraphics[trim={1em 0.5em 0.5em 0.5em},clip]{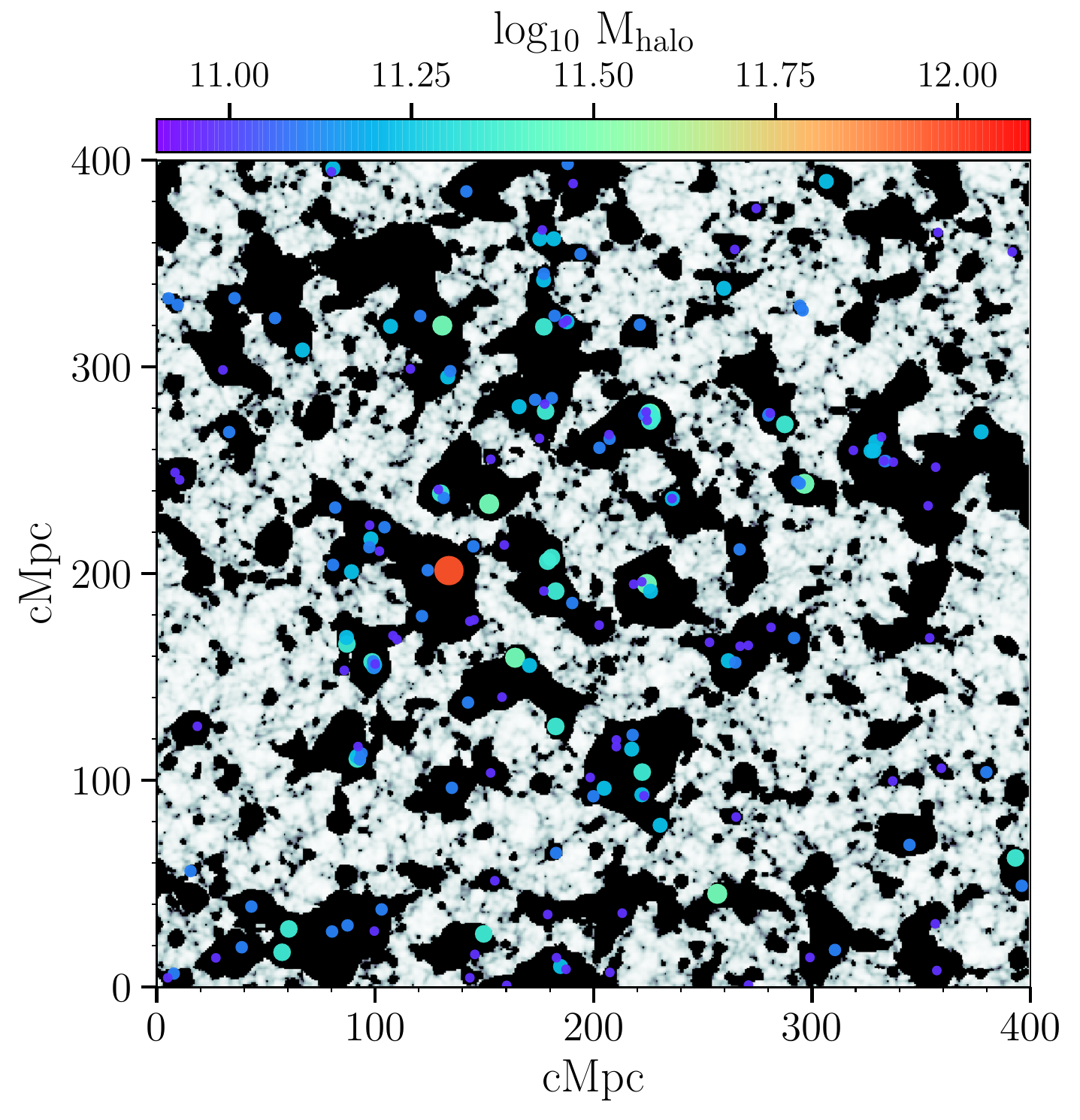}}\\
\end{center}
\caption{One pixel ($\sim800$ comoving kpc) slice through the semi-numerical ionization field at $z=7.5$ with $\langle x_{\rm HI}\rangle=0.5$ (greyscale; a linear stretch with white corresponding to neutral and black corresponding to fully ionized). The locations of $M_{\rm h}>10^{11}$ M$_\odot$ halos are shown from a 4 comoving Mpc-thick slice centered on the ionization field slice, color- and size-coded by halo mass. Massive halos tend to lie inside of large regions that have already been ionized.}
\label{fig:ionfield}
\end{figure}

Massive dark matter halos reside within larger-scale overdensities which are reionized early \citep{AA07}, leading to an important bias in the distribution of distances to the nearest patch of neutral gas \citep{Lidz07,MF08}.  We constructed dark matter halos directly from the initial conditions following the method of \citet{MF07} as now implemented in the public release of \texttt{21cmFAST}. In Figure~\ref{fig:ionfield}, we show a 0.78 Mpc-thick slice
through an ionization field at $z=7.5$ with
$\langle{x_{\rm HI}}\rangle=0.5$ and the locations of massive halos within $\pm2$ Mpc of the slice. As expected for the inside-out progression of reionization \citep{Furlanetto04}, halos preferentially (in fact, exclusively, for the massive halos shown here) lie inside of ionized regions (black).
As seen by the subtle grey shading,
the ionization field is not entirely a binary neutral (white) vs. ionized (black) field -- \texttt{21cmFAST} by default includes a prescription for a (typically small) degree of partial ionization due to ``unresolved" ionized bubbles within each cell, but this is unlikely to have any impact on our results.

\begin{figure}[htb]
\begin{center}
\resizebox{8.50cm}{!}{\includegraphics[trim={2.0em 0em 2.5em 3em},clip]{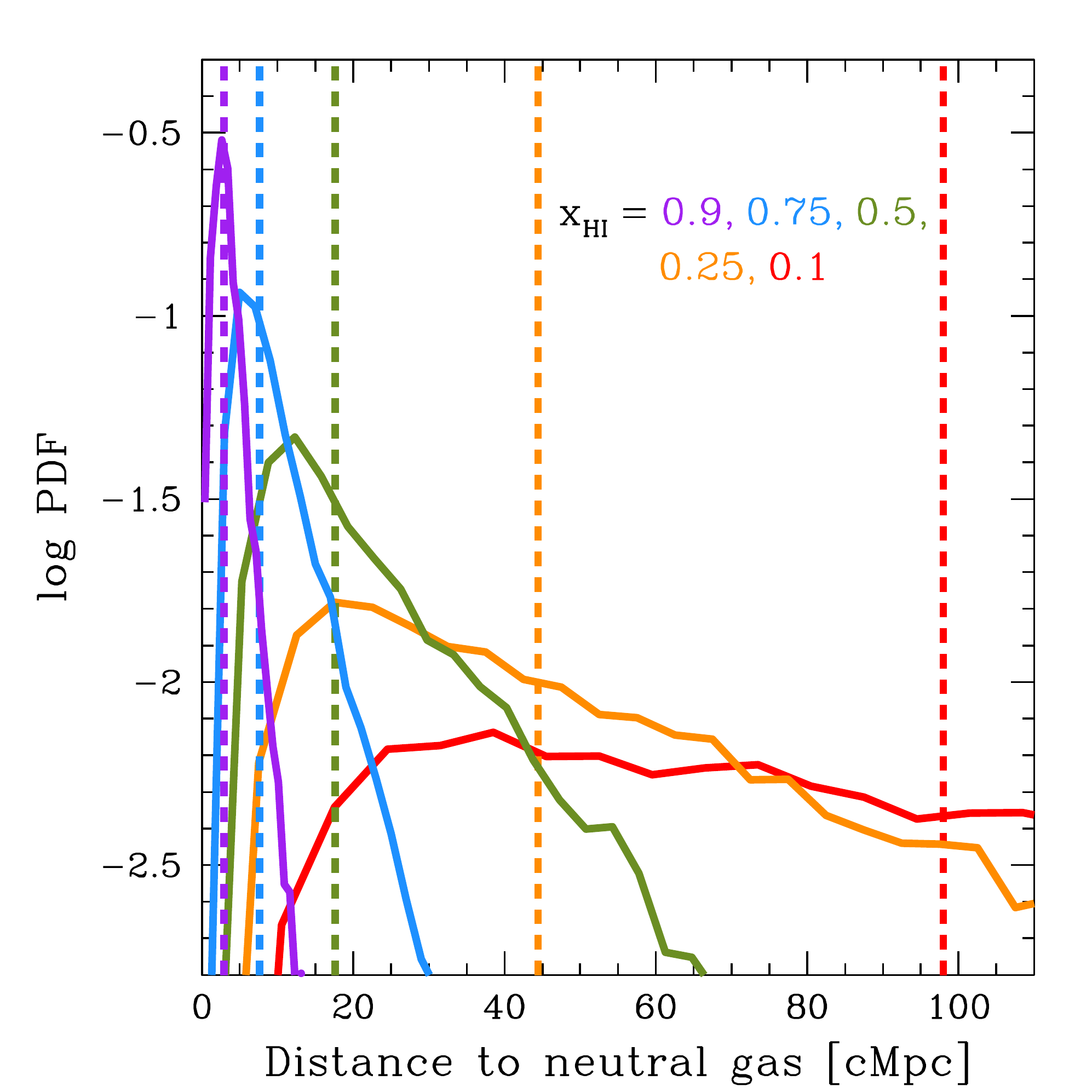}}\\
\end{center}
\caption{Distribution of distances from massive halos ($M_{\rm halo}\ga3\times10^{11}$ M$_\odot$) to the first patch of neutral gas in our semi-numerical simulations with global neutral fractions $\langle x_{\rm HI}\rangle=0.1,0.25,0.5,0.75,0.9$ from right to left. The vertical dashed lines correspond to the median of the correspondingly-colored distribution.}
\label{fig:bubbledist}
\end{figure}

We then extracted randomly-oriented sightlines of $x_{\rm HI}$ from the locations of the 500 most massive halos, corresponding to $M_{\rm h}\ga3\times10^{11}$ M$_\odot$. We show the distribution of distances from these massive halos
to the nearest patch of neutral hydrogen ($x_{\rm HI} > 0.01$) as a function of $\langle{x_{\rm HI}}\rangle$ in Figure~\ref{fig:bubbledist}.
This distance is what determines the initial strength of the damping wing feature when the quasar first turns on, and the distribution of distances leads to large variations in the damping wing profiles between different sightlines at the same global neutral fraction.
More massive halos tend to sit in larger ionized bubbles \citep{AA07,Lidz07,MF08}, so assuming a lower (higher) halo mass
than the actual quasar host halo would result in shorter (longer) distances to the nearest neutral patch at fixed $\langle{x_{\rm HI}}\rangle$, leading to an overestimate (underestimate) of the damping wing strength. In this work we ignore this potential source of bias, and note that it is likely degenerate with other assumptions in our model for the reionization topology (e.g. $M_{\rm min}$).

\subsection{Ionizing radiative transfer}

We use an updated version of the one-dimensional radiative transfer implementation in \citep{Davies16} to compute the effect of quasar radiation on the surrounding IGM, which we briefly summarize below. The radiative transfer code computes the time-dependent evolution of six species (e$^{-}$, \ion{H}{1}, \ion{H}{2}, \ion{He}{1}, \ion{He}{2}, \ion{He}{3}) and the gas temperature, following the method described in the Appendix of \citet{BH07}. The abundances of ionic species are computed by integrating the following
coupled system of equations,
\begin{eqnarray}
\frac{dn_\HII}{dt} &=& n_\HI (\Gamma^\gamma_{\HI} + n_e \Gamma^\mathrm{e}_{\HI}) - n_\HII n_e \alpha^A_\HII, \\
\frac{dn_\HeII}{dt} &=& n_\HeI (\Gamma^\gamma_{\HeI} + n_e \Gamma^\mathrm{e}_{\HeI}) + n_\HeIII n_e \alpha^A_\HeIII \nonumber \\
&& - n_\HeII (\Gamma^\gamma_{\HeII} + n_e \Gamma^\mathrm{e}_\HeII - n_e \alpha^A_\HeII), \\
\frac{dn_\HeIII}{dt} &=& n_\HeII (\Gamma^\gamma_{\HeII}  + n_e \Gamma^\mathrm{e}_\HeII) - n_\HeIII n_e \alpha^A_\HeIII,
\end{eqnarray}
where $n_i$ are the number densities for each species, $\Gamma^\gamma_i$ are the photoionization rates, $\Gamma^\mathrm{e}_i$ are the collisional ionization rates, and $\alpha^A_i$ are the Case A recombination rate coefficients. In $\Gamma^{\gamma}_i$ we include the effect of \emph{secondary} ionizations, as tabulated by \citet{FJS10}, whereby energetic photoelectrons (with kinetic energy greater than the ionization potential) lose energy by ionizing additional atoms rather than simply dumping the excess photoionization energy into the gas as heat. The remaining species are then solved for via the closure
conditions
\begin{eqnarray}
n_\HI &=& n_\mathrm{H} - n_\HII, \\
n_\HeI &=& \frac{Y}{4(1-Y)}n_\mathrm{H} - n_\HeII - n_\HeIII, \\
n_e &=& n_\HII + n_\HeII + 2n_\HeIII,
\end{eqnarray}
where we have assumed $Y=0.24$ for the mass fraction of helium.

The gas temperature is evolved taking into account photoionization heating and cooling\footnote{We assume that the gas is of primordial composition, i.e. there is no cooling due to elements heavier than helium.} from recombinations, collisional excitation, the expansion of the Universe, and inverse Compton scattering off of
CMB photons (see \citealt{Davies16} for more details). Adding to the model presented in \citet{Davies16}, we now include the prescription from \citet{Rahmati13} to approximate self-shielding of the ionizing background in dense gas, and update this self-shielding at each time step to take into account the ionization of dense absorbers by the quasar.

For the ionizing spectrum of each quasar, we first use the \citet{Lusso15} template to convert from the measured $M_{1450}$ to $L_\nu$ at the ionizing edge of hydrogen ($E_{\rm HI}\approx13.6$ eV), and then extrapolate to higher frequencies by assuming $L_\nu \propto \nu^{-1.7}$, in agreement with the best-fit power-law spectrum from \citet{Lusso15}. Assuming a different average quasar template (e.g. \citealt{Telfer02,Stevans14}) could change the output of ionizing photons by tens of percent, which would have implications for the shape of the proximity zone transmission profile (Davies et al., in prep.) and strength of the damping wing. However, the size of the ionized bubble around the quasar $R_{\rm ion}$ (assuming a fully neutral universe) is only weakly dependent on the ionizing photon output $\dot{N}_{\rm ion}$ ($R_{\rm ion} \propto \dot{N}_{\rm ion}^{1/3}$; \citealt{CH00}), and this dependence is completely degenerate with the lifetime of the quasar.
When computing the photoionization and photoheating rates, we integrate the ionizing spectrum over frequencies from the ionizing frequency $\nu_i$ to $40\nu_i$ for each (partially-)neutral species $i$, separately, with 25 logarithmic frequency bins.

We assume a ``lightbulb" model for quasar emission: the quasar turned on at some point $t_{\rm q}$ in the past,
and has been shining at a constant luminosity since then. In the rest of the paper we will refer to $t_{\rm q}$ as the ``quasar lifetime."

\subsection{Hybrid model}\label{sec:model}

\begin{figure*}[htb]
\begin{center}
\resizebox{17.6cm}{!}{\includegraphics[trim={7.0em 2em 6.5em 2.5em},clip]{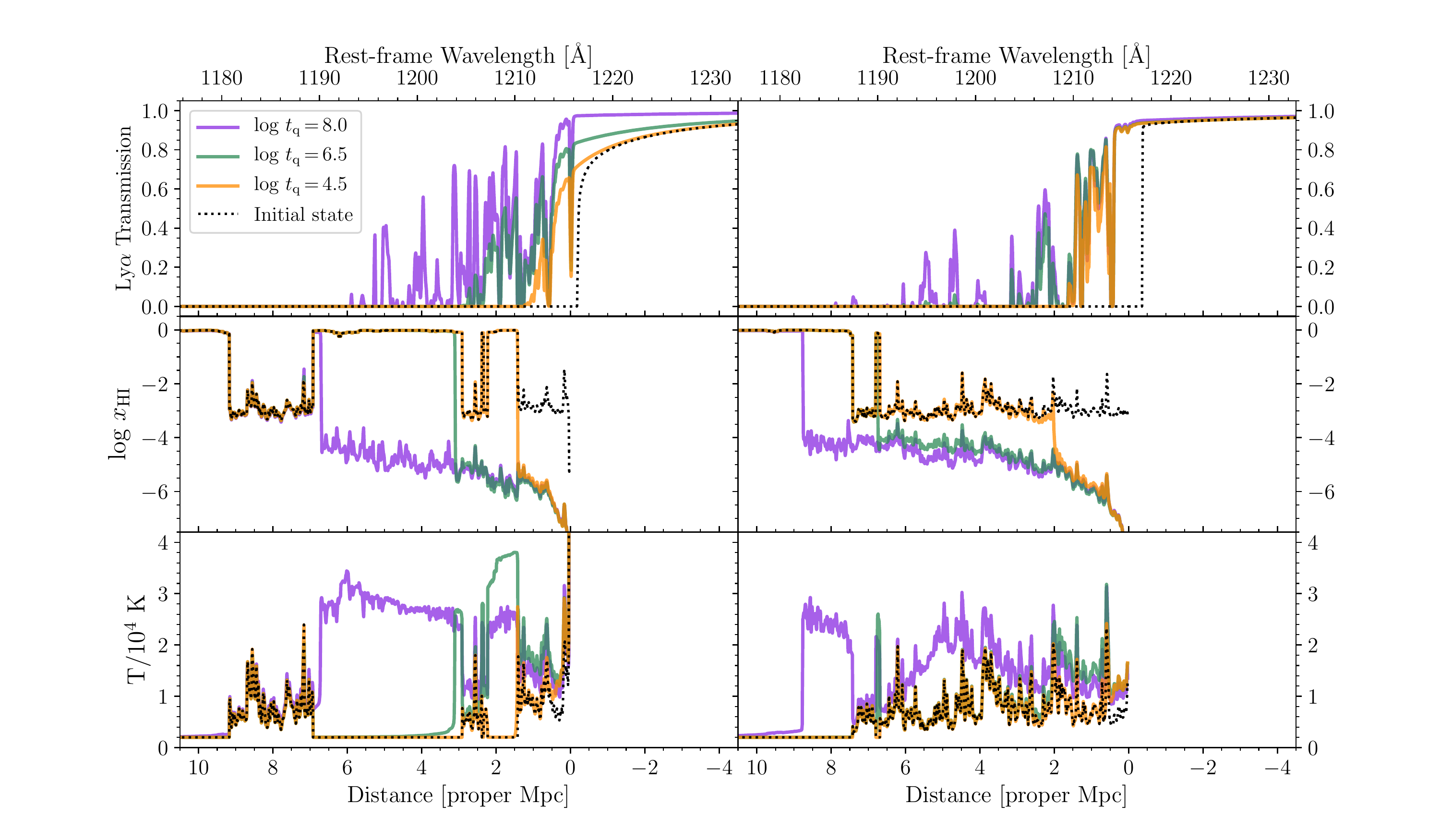}}\\
\end{center}
\caption{Example outputs from the hybrid model of quasar proximity zones at $\log{t_{\rm q}}=$ 4.5 (orange), 6.5 (green), and 8.0 (purple) for two skewers (left and right) through the $\langle x_{\rm HI}\rangle=0.5$ simulation at $z=7.54$. The black dotted curves show the initial state of the skewer prior to the quasar turning on. The top panels show the Ly$\alpha$ transmission, the middle panels show $x_{\rm HI}$, and the bottom panels show the gas temperature.}
\label{fig:model_ex}
\end{figure*}

\begin{figure}[htb]
\begin{center}
\resizebox{8.50cm}{!}{\includegraphics[trim={1.3em 2em 4.0em 2.5em},clip]{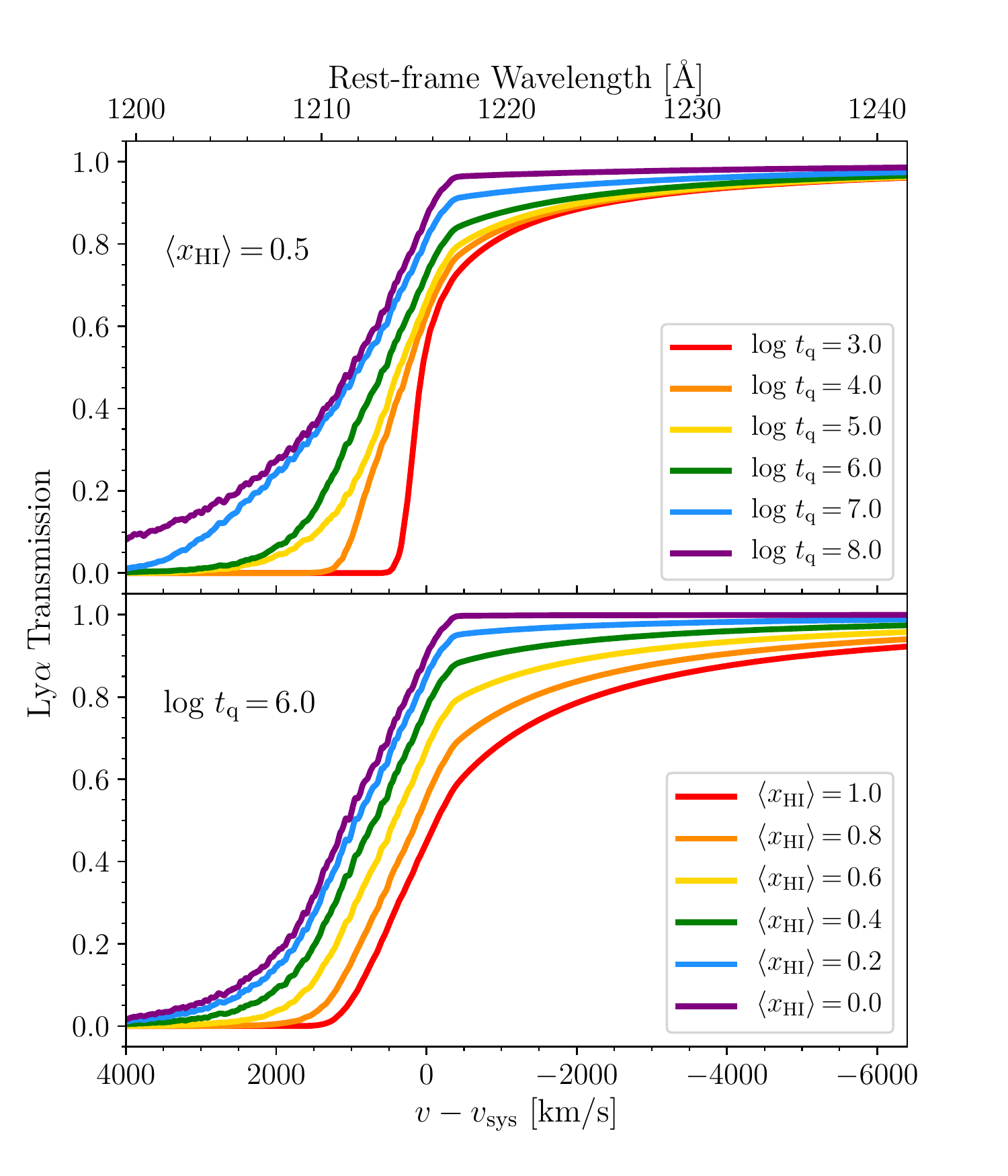}}\\
\end{center}
\caption{Mean Ly$\alpha$ transmission profiles from the 1D radiative transfer simulations. The top panel shows the variation in the mean profile for varying quasar lifetime ($\log{[t_{\rm q}/{\rm yr}]}$=3.0--8.0, $\Delta\log{t_{\rm q}}$=1.0) at a fixed global neutral fraction $\langle x_{\rm HI}\rangle$=0.5. The bottom panel shows the variation in the mean profile for varying neutral fraction ($\langle x_{\rm HI}\rangle$=0.0--1.0, ${\Delta}x_{\rm HI}$=0.2) at a fixed quasar lifetime of $\log{[t_{\rm q}/{\rm yr}]}=6.0$.}
\label{fig:model_stacks}
\end{figure}

We synthesize these three model components by computing ionizing radiative transfer along hydrodynamical simulation skewers, with the initial neutral fraction along each sightline set by skewers from the semi-numerical reionization simulations. For regions with $x_{\rm HI}=0$ in the reionization simulations, we assume a uniform ionizing background such that $\langle x_{\rm HI}\rangle$ inside of ionized regions is $\sim10^{-3}$ (corresponding to a hydrogen photoionization rate $\Gamma_{\rm HI}\sim6\times10^{-14}$), although we find that our results are insensitive to this choice. 
We initialize the IGM temperature in ionized regions to the values from the hydrodynamical simulation skewer, and assume that the IGM is initially cold (2000 K) inside of neutral regions (e.g. \citealt{Furlanetto06}).
The density field from the hydrodynamical simulation does not reflect the additional clumpiness of such cold gas, i.e. the gas in the simulation has been ``pressure smoothed"
to some extent \citep{GH98,Rorai13,Kulkarni15}, but we do not expect this to have a large effect on the transmission profile. 

In the left column of Figure~\ref{fig:model_ex}, we show a typical example of the output from our full model, a simulated sightline assuming the luminosity and redshift of J1342+0928 ($M_{1450}=-26.76, z_{\rm q}=7.5413$) using a skewer from the hydrodynamical simulation with initial $x_{\rm HI}$ given by a skewer from the $\langle x_{\rm HI}\rangle=0.5$ semi-numerical reionization simulation.
The top panel shows the Ly$\alpha$ transmission in the quasar spectrum, the middle panel shows the neutral fraction, and the bottom panel shows the gas temperature. The different colors correspond to quasar lifetimes of $10^{4.5}$ (orange), $10^{6.5}$ (green), and $10^8$ (purple) years. The damping wing signal, shown by the absorption at negative distance ($\lambda_{\rm rest}>\lambda_{{\rm Ly}\alpha}$), is very strong soon after the quasar turns on (orange), with the first patch of neutral gas encountered at $\sim1.3$ proper Mpc along the line of sight.
This first neutral patch is ionized within a few million years (green), weakening the damping wing considerably and photoheating the gas to $T\sim3$--$4\times10^4$~K.
After 100 million years (purple), the quasar has carved out a large enough
ionized region ($\sim7$ proper Mpc) to completely wipe out the damping wing signal, and initially-neutral photoheated regions near the quasar have cooled substantially.
At this stage, the proximity zone is no longer cut off by the onset of fully neutral gas in the IGM; instead, the Ly$\alpha$ forest absorption becomes too strong as the ionizing flux from the quasar decreases (e.g. \citealt{BH07,Eilers17}) and
$x_{\rm HI}$ reaches $\sim10^{-4}$, as shown by the complete disappearance of the purple transmission curve at a shorter distance ($\la 6$ proper Mpc)
than the location
of the
ionization front (6.7 proper Mpc).

Cosmic variance in the density field and in the reionization morphology lead to a wide variety of proximity zone and damping wing spectra at the same neutral fraction -- we show another example sightline in the right column of Figure~\ref{fig:model_ex} which initially resides in a very large ionized region, and thus \emph{never} shows a strong damping wing signal.
For the longest lifetime model (purple), a modest amount of extra transmission appears at
$R\sim5$ proper Mpc
due to heating from the reionization of \ion{He}{2} by the quasar (the thermal proximity effect, e.g. \citealt{Bolton12,Khrykin17}).

For each semi-numerical reionization box, corresponding to $0 \leq \langle x_{\rm HI}\rangle \leq 1$ in
21 steps of $\Delta x_{\rm HI}=0.05$, we ran 2400 radiative transfer simulations on 2400 different random skewers, using each of our 1200 hydrodynamical simulation skewers twice. From these simulations, we computed transmission spectra every
$\Delta \log{[t_{\rm q}/yr]} = 0.5$
in 11 steps from $10^3$ to $10^8$ years covering a velocity range $-10,000 \leq v-v_{\rm sys} \leq +10,000$ km/s, where $v_{\rm sys}$ is the systemic velocity of the halo center. Our final set of transmission spectrum models is then $21 \times 11$, with 2400 spectra for each point in the coarse 2D grid. In Figure~\ref{fig:model_stacks} we show the mean transmission profiles from our simulations as a function of $t_{\rm q}$ at fixed $\langle x_{\rm HI}\rangle=0.5$ (top), and as a function of $\langle x_{\rm HI}\rangle$ at fixed $t_{\rm q}=10^6$ years (bottom).
There is a clear trend towards stronger damping wings and smaller proximity zones for high neutral fractions and short quasar lifetimes.

As noted by \citet{Bolton11}, a degeneracy exists between IGM neutral fraction and quasar lifetime in determining the shape of the proximity zone and damping wing profile, wherein short quasar lifetime and small neutral fraction appears similar to long quasar lifetime and large neutral fraction (
although these models assumed a constant $x_{\rm HI}$ in the IGM instead of our
more realistic patchy topology).
A similar degeneracy arises in our hybrid model because even at large neutral fraction, the
quasar can carve out a large ionized region that greatly increases the distance to the nearest neutral patch,
decreasing the strength of the damping wing feature and increasing the size of the proximity zone (see the purple curves in Figure~\ref{fig:model_ex}). Another consequence of this is that at relatively long quasar lifetimes, $t_{\rm q}\ga10^8$ years, the damping wing almost entirely disappears, even for $\langle x_{\rm HI}\rangle\sim1$.

At even shorter timescales, $t_{\rm q}\la10^5$ years, the inner parts of the proximity zone start to disappear entirely (see the orange curves in Figure~\ref{fig:model_ex}).
This occurs because the gas has not been illuminated long enough
to respond to the increased ionizing flux from the quasar (e.g. \citealt{Khrykin16}), and such short lifetimes may explain the handful of very small proximity zones observed at $z\sim6$ (\citealt{Eilers17}, Davies et al. in prep.).

\section{Statistical Method for Jointly Inferring the Neutral Fraction and Quasar Lifetime}\label{sec:stats}

The measured quasar proximity zone and damping wing signals are a highly covariant heteroskedastic process, with large sightline-to-sightline variance for any particular set of parameters ($\langle x_{\rm HI}\rangle,t_{\rm q}$).
In addition to the 
uncorrelated
photon noise in the spectrum, 
additional sources of variance are the IGM density field, which leads to the
absorption inside the proximity zone, and the distance to the nearest neutral patch of the IGM, which has strong covariant effects across the whole spectrum. The uncertainty in our prediction for the quasar continuum (\S~2) introduces an additional multiplicative error which is strongly covariant. The combination of these processes cannot be simply described by
by a multivariate Gaussian likelihood, suggesting that inference via standard likelihood-based methods (e.g. Markov Chain Monte Carlo) may be difficult to interpret correctly.

Instead, we adopt an approach following principles of Bayesian indirect inference \citep{Gourieroux93,Drovandi15}, wherein the likelihood for auxiliary parameters or an auxiliary likelihood of the true parameters is used in place of an intractable true likelihood for the true parameters.
We define a ``pseudo-likelihood" $\tilde{L}$ as the product
of flux
probability distribution functions (PDFs)
$P(F_i)$ of 500 km/s binned pixels,
\begin{equation}\label{eqn:pseudo}
\tilde{L}(\theta) = \prod_i P(F_i|\theta),
\end{equation}
which is equivalent to the likelihood function of the (500 km/s binned) transmission spectrum in the absence of correlations between pixels.
For computational simplicity, and to limit the impact of our finite number of simulated sightlines, we approximate the flux PDFs $P(F_i|\theta)$ of each bin $i$ by fitting them with mixtures of three Gaussians\footnote{The exact form of the approximation to the individual flux PDFs appears to have only a minor effect on our analysis -- similar, albeit somewhat less constraining, posterior PDFs can be obtained with single Gaussian fits.}. While direct parameter inference from this likelihood would be formally incorrect due to the neglected correlations, one can still determine a set of maximum pseudo-likelihood model parameters, $\theta_{\rm M\tilde{L}E}$, which should be closely related to the true maximum likelihood parameters. This procedure
reduces the dimensionality of our data from the number of transmitted flux bins in the spectrum to the number of model parameters, allowing for a full Bayesian treatment with modest computational expense (albeit likely with slightly
less constraining power than the original data due to information lost in this compression).

We treat $\theta_{\rm M\tilde{L}E}$ as a summary statistic and compute the posterior PDF of the ``true" model parameters $\theta$ following Bayes' theorem,
\begin{equation}
p(\theta|\theta_{\rm M\tilde{L}E}) = \frac{p(\theta_{\rm M\tilde{L}E}|\theta)p(\theta)}{p(\theta_{\rm M\tilde{L}E})},
\end{equation}
where $p(\theta|\theta_{\rm M\tilde{L}E})$ is the posterior PDF of $\theta$, $p(\theta_{\rm M\tilde{L}E}|\theta)$ is the likelihood of $\theta_{\rm M\tilde{L}E}$ given the model $\theta$,
 $p(\theta)$ is the prior on $\theta$, and $p(\theta_{\rm M\tilde{L}E})$ is the evidence.
We compute the likelihood function \emph{directly} by measuring the distribution of
$\theta_{\rm M\tilde{L}E}$ for forward-modeled mock data on a coarse grid of $\theta=(\langle x_{\rm HI}\rangle,t_{\rm q})$ and explicitly computing the evidence,
\begin{equation}
p(\theta_{\rm M\tilde{L}E})=\int p(\theta_{\rm M\tilde{L}E}|\theta) p(\theta) d\theta.
\end{equation}
We denote as $\hat{F}$ a forward-modeled transmission spectrum, which results from taking a draw $F$ from the set of 2400 transmission spectra computed for the parameter set $\theta$,
multiplication by a random draw from a multivariate Gaussian distribution describing the relative continuum error ($\epsilon_C$; Paper I), then a draw of additive Gaussian noise following the continuum-normalized noise vector of the spectrum ($N$):
\begin{equation}
\hat{F} = F\times(1+\epsilon_C)+N.
\end{equation}
We find the $\theta_{\rm M\tilde{L}E}$ for each mock spectrum via a simple brute force approach, computing $\tilde{L}$ (equation \ref{eqn:pseudo})
for each of the $21\times11$ models in our coarse grid.
For priors, we assume a flat \emph{linear} prior on $\langle x_{\rm HI}\rangle$ from 0 to 1, and flat \emph{log} priors on $t_{\rm q}$ from $10^{3}$ to $10^{8}$ years. The linear prior on $\langle x_{\rm HI}\rangle$ reflects our expectation that $z>7$ is in the midst of the reionization epoch, while the log prior on $t_{\rm q}$ reflects our broad uncertainty on the lifetime of the luminous quasar phase, incorporating recently discovered evidence for very short lifetimes \citep{Eilers17}. We will also quote constraints for a stronger prior on $t_{\rm q}$ which only allows lifetimes as short as $10^5$ years. We will discuss the impact of this choice of priors in \S~\ref{sec:priors}.

In this method, the data have been compressed into the measurement of $\theta_{\rm M\tilde{L}E}$, so any spectrum with a particular $\theta_{\rm M\tilde{L}E}$ will result in an identical 2D posterior PDF. Given our coarse grid in parameter space, there are only $21\times11=231$ possible posterior PDFs for any given quasar spectrum.

\section{Results} % PCA results + posterior PDFs of xhi + tq

In Paper I, we estimated the intrinsic continua (and their uncertainties) of the two quasars known at $z>7$: ULAS J1120+0641 ($z=7.0851$; \citealt{Mortlock11,Venemans17}) and ULAS J1342+0928 ($z=7.5413$; \citealt{Banados18,Venemans17b}). Here we apply the statistical method from the previous section to jointly constrain the neutral fraction at $z>7$ and the quasar lifetimes by comparing the resulting transmission profiles to our simulated spectra.

For the purposes of modeling the physical state of the IGM along the line of sight, we adopt the precise systemic redshifts above as the true locations of the quasar host halos. However, these systemic redshifts have little relevance to the PCA continua, given that the PCA model was trained on quasars with imprecise redshifts. In Paper I we resolved this ambiguity by fitting for a ``template redshift" simultaneously with the red-side PCA coefficients, resulting in an independent (but physically irrelevant) redshift estimate that can be applied to any quasar spectrum. This template redshift is what then defines the rest-frame wavelengths for the continuum prediction. 

\subsection{ULAS J1120+0641}

\begin{figure*}[ht]
\begin{center}
\resizebox{17.6cm}{!}{\includegraphics[trim={6.5em 0em 8.0em 4.0em},clip]{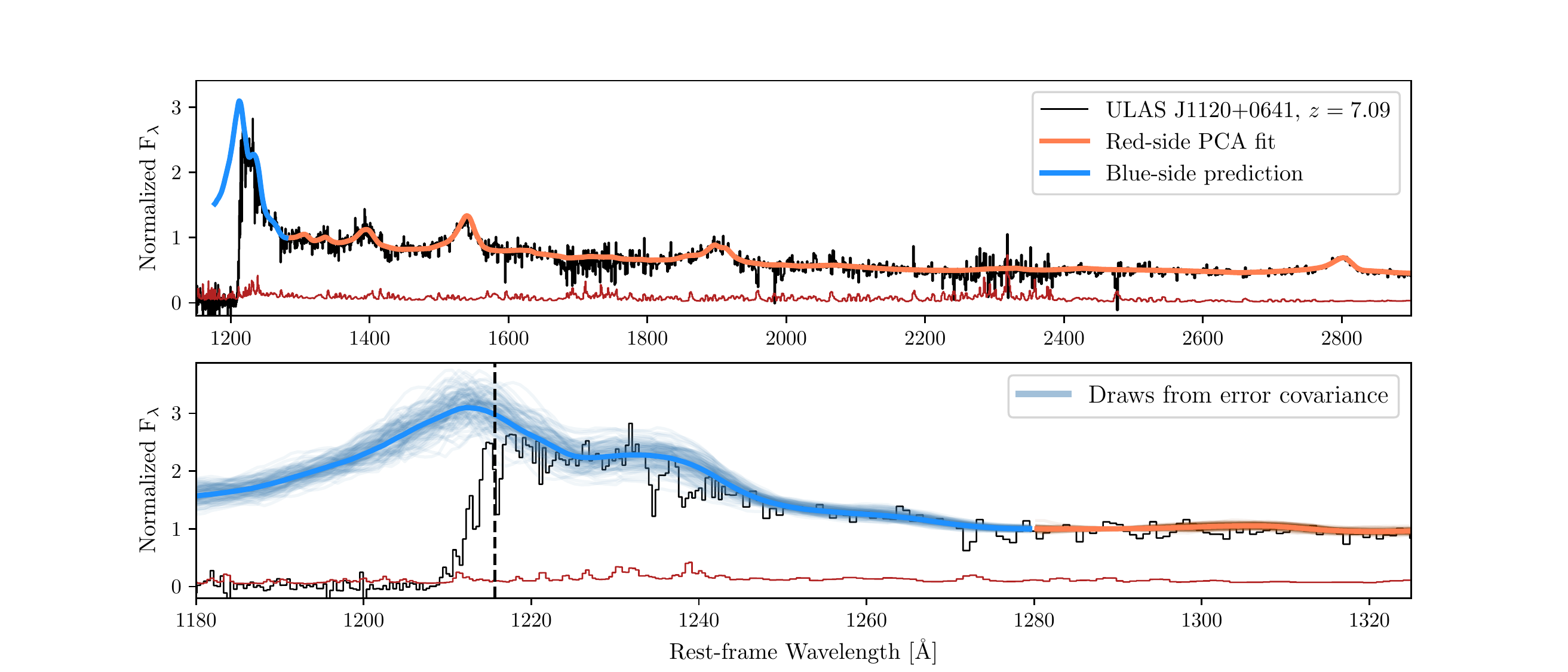}}\\
\end{center}
\caption{Top: VLT/FORS2 + Gemini/GNIRS spectrum of ULAS J1120+0641 (\citealt{Mortlock11}, black) and its noise vector (red). The red-side PCA fit and blue-side prediction are shown as the orange and blue curves, respectively. Bottom: Zoom in of the Ly$\alpha$ region of the spectrum, where the vertical dashed line shows rest-frame Ly$\alpha$ ($\lambda_{\rm rest}=1215.67$ {\AA}). The transparent curves show 100 draws from the covariant blue-side prediction error calibrated from the 1\% most similar quasars in the training set. This quasar shows modest evidence for a damping wing and has a relatively small proximity zone.}
\label{fig:mortlock_pca}
\end{figure*}

\begin{figure}[htb]
\begin{center}
\resizebox{8.5cm}{!}{\includegraphics[trim={1.2em 1.2em 1.2em 1em},clip]{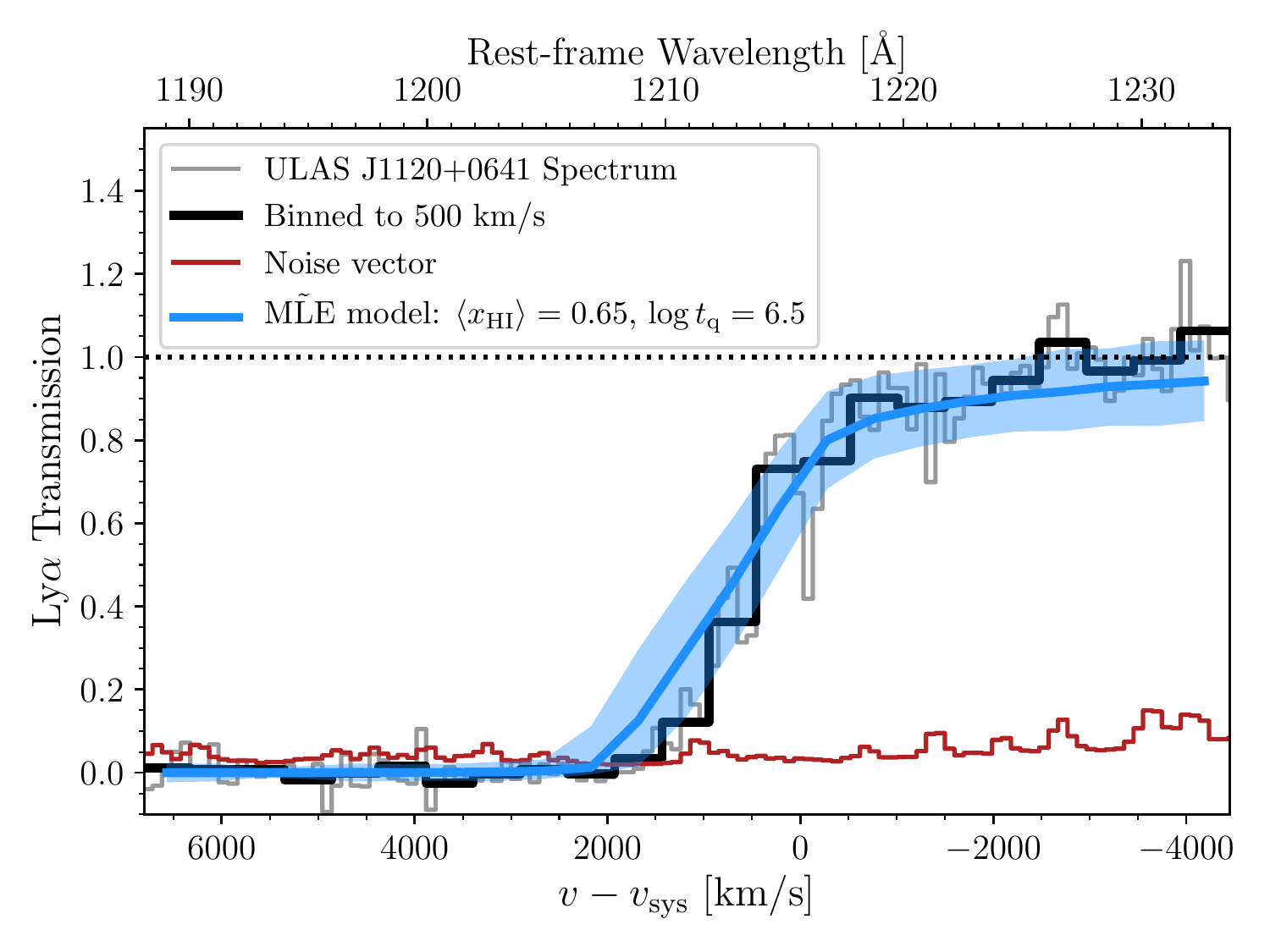}}\\
\end{center}
\caption{Continuum-divided spectrum of ULAS J1120+0641 (grey) and its noise vector (red). The black histogram shows the spectrum rebinned to $\sim500$ km/s in the region we use for model comparison. The blue solid curve shows the median $\sim500$ km/s-binned transmission spectrum of mock spectra with the M\~{L}E parameter values $\theta_{\rm M\tilde{L}E}=(\langle x_{\rm HI}\rangle=0.65,\log{t_{\rm q}}=6.5)$, while the associated blue shaded region shows the 16th--84th percentile range for mock spectra with $\theta=\theta_{\rm M\tilde{L}E}$}
\label{fig:mortlock_fit}
\end{figure}

\begin{figure}[htb]
\begin{center}
\resizebox{8.50cm}{!}{\includegraphics[trim={6.0em 0.2em 4.0em 3.5em},clip]{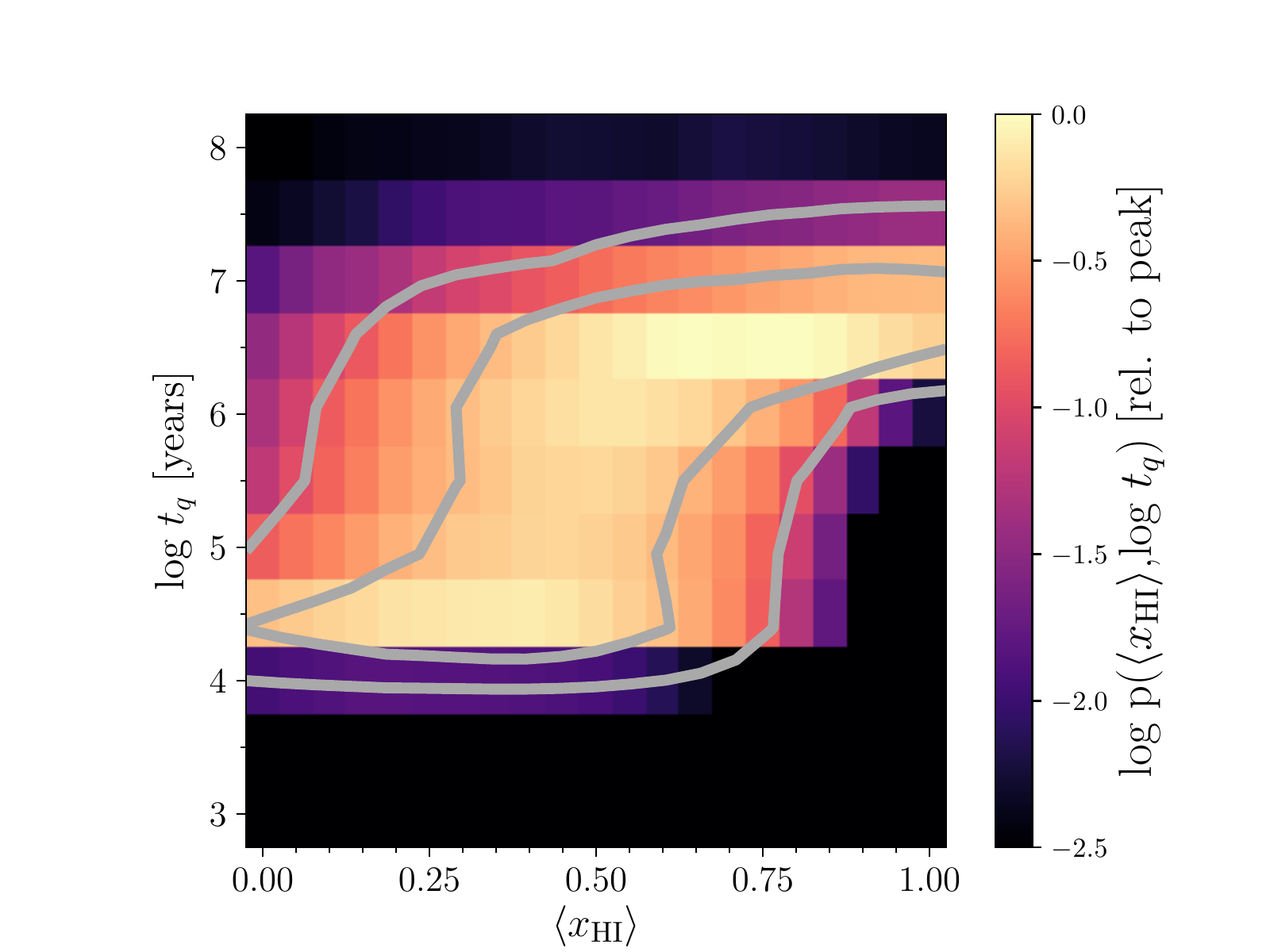}}\\
\end{center}
\caption{2D posterior PDF of $\langle x_{\rm HI}\rangle$ and $\log{t_{\rm q}}$ resulting from the M\~{L}E parameter values $\theta_{\rm M\tilde{L}E}=(\langle x_{\rm HI}\rangle=0.65,\log{t_{\rm q}}=6.5)$ derived from the ULAS J1120+0641 spectrum. The contours enclose $68\%$ and $95\%$ of the total probability.}
\label{fig:mortlock_post}
\end{figure}

In Figure~\ref{fig:mortlock_pca}, we show the red-side fit to the VLT/FORS2 + Gemini/GNIRS spectrum of ULAS J1120+0641 (\citealt{Mortlock11}; top panel) and the predicted blue-side
continuum (bottom panel) from Paper I. The predicted continuum has been corrected for the mean bias of predicted continua for similar quasars in the training set, as discussed in \S\ref{sec:pca}
and shown in Figure 12 of Paper I. We find a best-fit template redshift of $z=7.0834$,
a very small blueshift of ${\Delta}v=63$ km/s
from the systemic
frame defined by the centroid of the [CII] emission line of the host galaxy ($z=7.0851$, \citealt{Venemans17}). 
The blue-side profile shows a hint of absorption redward of Ly$\alpha$ and a relatively small proximity zone ($R_{\rm p}=1.72$ proper Mpc,
following the definition in \citealt{Eilers17}) compared to the trend seen at $z\sim5.7$--$6.5$ \citep{Eilers17,Mazzucchelli17}.
While the damping wing signal appears to be fairly strong at rest-frame Ly$\alpha$, similar
to
what was found by previous works \citep{Mortlock11,Simcoe12,Greig17b},
when our large, covariant continuum uncertainty is taken into account the spectrum does not appear to definitively indicate a neutral IGM. 

We show the resulting transmission spectrum (i.e. observed spectrum divided by the continuum model) as the grey curve in Figure~\ref{fig:mortlock_fit} and the 500 km/s-binned spectrum in black. For the statistical analysis,
we only use pixels at $v-v_{\rm sys} > -4,400$ km/s ($\lambda_{\rm rest}\la1233$ {\AA})
to avoid the strong (and unresolved) associated \ion{N}{5} absorption, and we choose to end the blue-side coverage at $v-v_{\rm sys} = +6,400$ km/s ($\lambda_{\rm rest}\sim1190$ {\AA})
because all of the proximity zone models have no detectable signal beyond that distance. We find maximum pseudo-likelihood parameter values of $\theta_{\rm M\tilde{L}{E}} = (\langle x_{\rm HI}\rangle=0.65, \log{t_{\rm q}}=6.5)$, and we show the median transmission profile of the M\~{L}E model (blue solid) and the expected 16--84th percentile scatter (blue shaded)
from forward modeled spectra with $\theta=\theta_{\rm M\tilde{L}E}$ in Figure~\ref{fig:mortlock_fit}.
From the M\~{L}E parameter values we infer the 2D posterior PDF $p(\theta|\theta_{\rm M\tilde{L}E})$ shown in Figure~\ref{fig:mortlock_post}. The posterior PDF is relatively flat across a wide swathe of $(\langle x_{\rm HI} \rangle,\log{t_{\rm q}})$ parameter space, with a trend towards higher $\langle x_{\rm HI} \rangle$ for longer $t_{\rm q}$, reflectingv the degeneracy between these two parameters discussed in \S~\ref{sec:model} and shown in Figure~\ref{fig:model_stacks}.
The non-zero size of the proximity zone rules out quasar lifetimes shorter than $\sim10^{4.5}$ years, while the combination of damping wing strength and small proximity zone rule out quasar lifetimes longer than $\sim10^{7}$ years.

\subsection{ULAS J1342+0928}

\begin{figure*}[htb]
\begin{center}
\resizebox{17.6cm}{!}{\includegraphics[trim={6.2em 0em 8.0em 4.0em},clip]{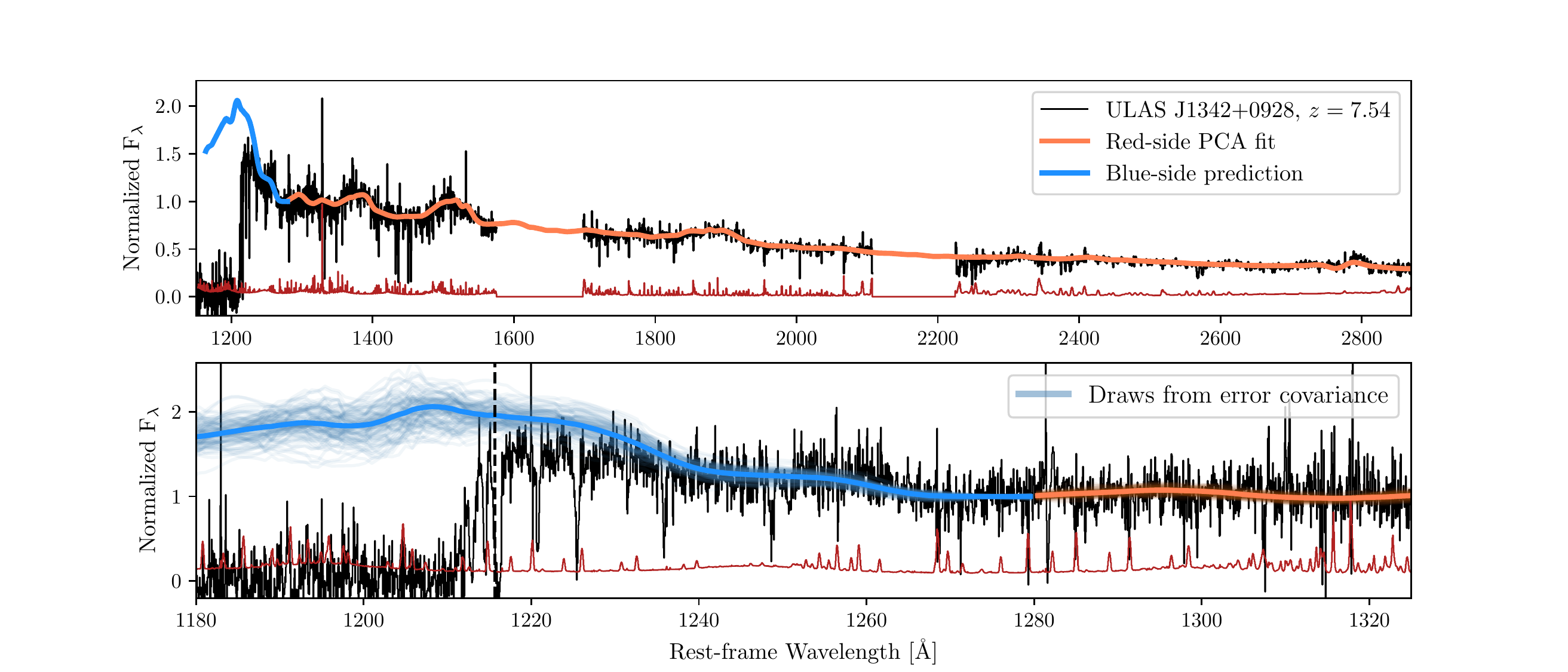}}\\
\end{center}
\caption{Similar to Figure~\ref{fig:mortlock_pca} but for the Magellan/FIRE + Gemini/GNIRS spectrum of ULAS J1342+0928 (black), including its noise vector (red), red-side PCA fit (orange), and blue-side prediction (blue). The FIRE spectrum in the top panel has been re-binned to match the resolution of the GNIRS data used in the K-band, while the bottom panel is shown at the higher FIRE resolution. This quasar shows strong evidence for a damping wing and has a very small proximity zone.}
\label{fig:pisco_pca}
\end{figure*}

\begin{figure}[htb]
\begin{center}
\resizebox{8.5cm}{!}{\includegraphics[trim={1.2em 1.2em 1.2em 0.2em},clip]{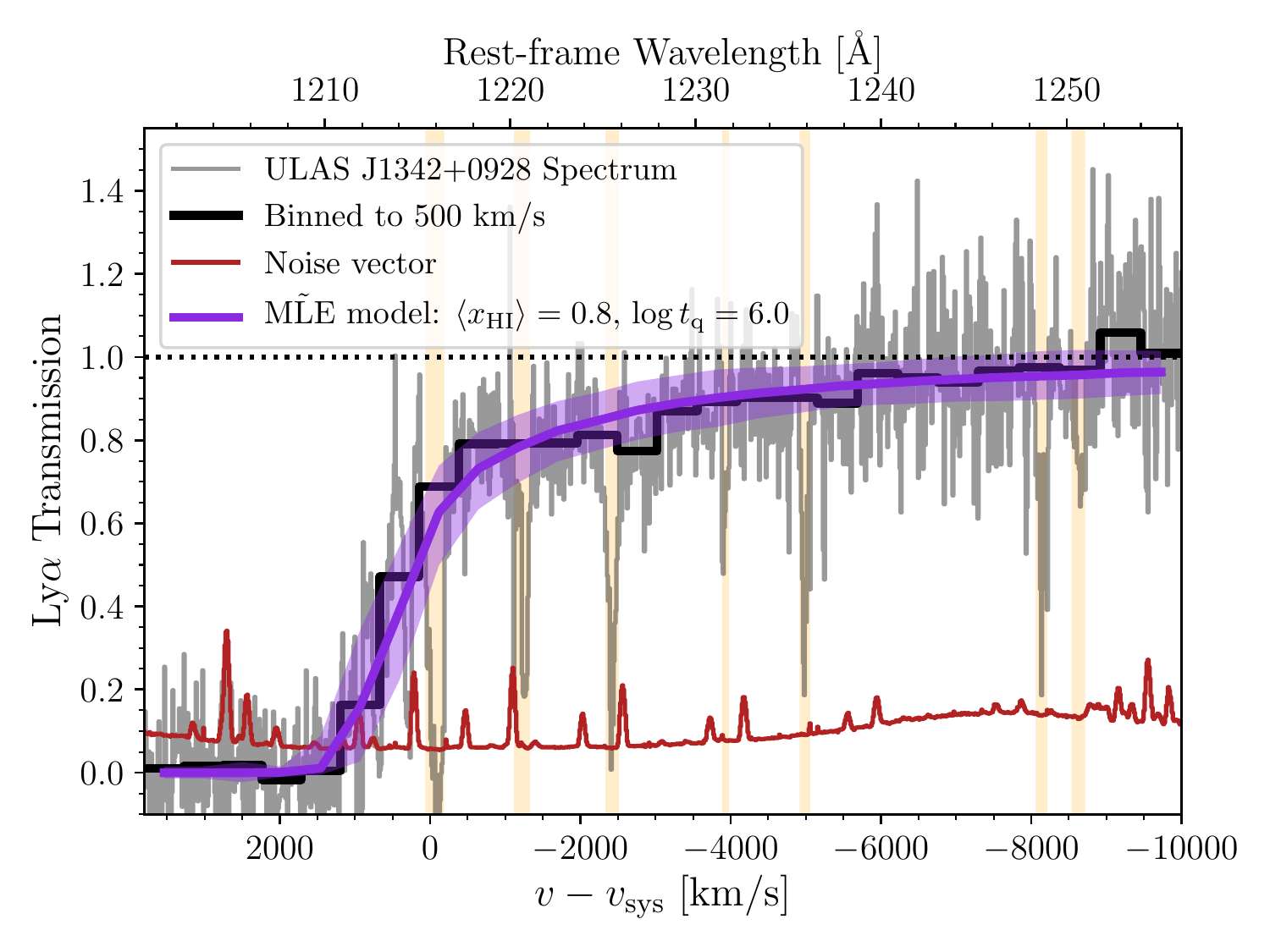}}\\
\end{center}
\caption{Similar to Figure~\ref{fig:mortlock_fit} but for the continuum-divided spectrum of ULAS J1342+0928. The purple solid curve shows the median binned transmission spectrum in the mock spectra assuming the M\~{L}E parameter values $\theta_{\rm M\tilde{L}E}=(\langle x_{\rm HI}\rangle=0.8,\log{t_{\rm q}}=6.0)$, while the associated purple shaded region shows the 16th--84th percentile range for mock spectra with $\theta=\theta_{\rm M\tilde{L}E}$. The orange shaded regions highlight identified metal absorption systems that we have masked in our analysis.}
\label{fig:pisco_fit}
\end{figure}

\begin{figure}[htb]
\begin{center}
\resizebox{8.50cm}{!}{\includegraphics[trim={6.0em 0.0em 4.0em 3.5em},clip]{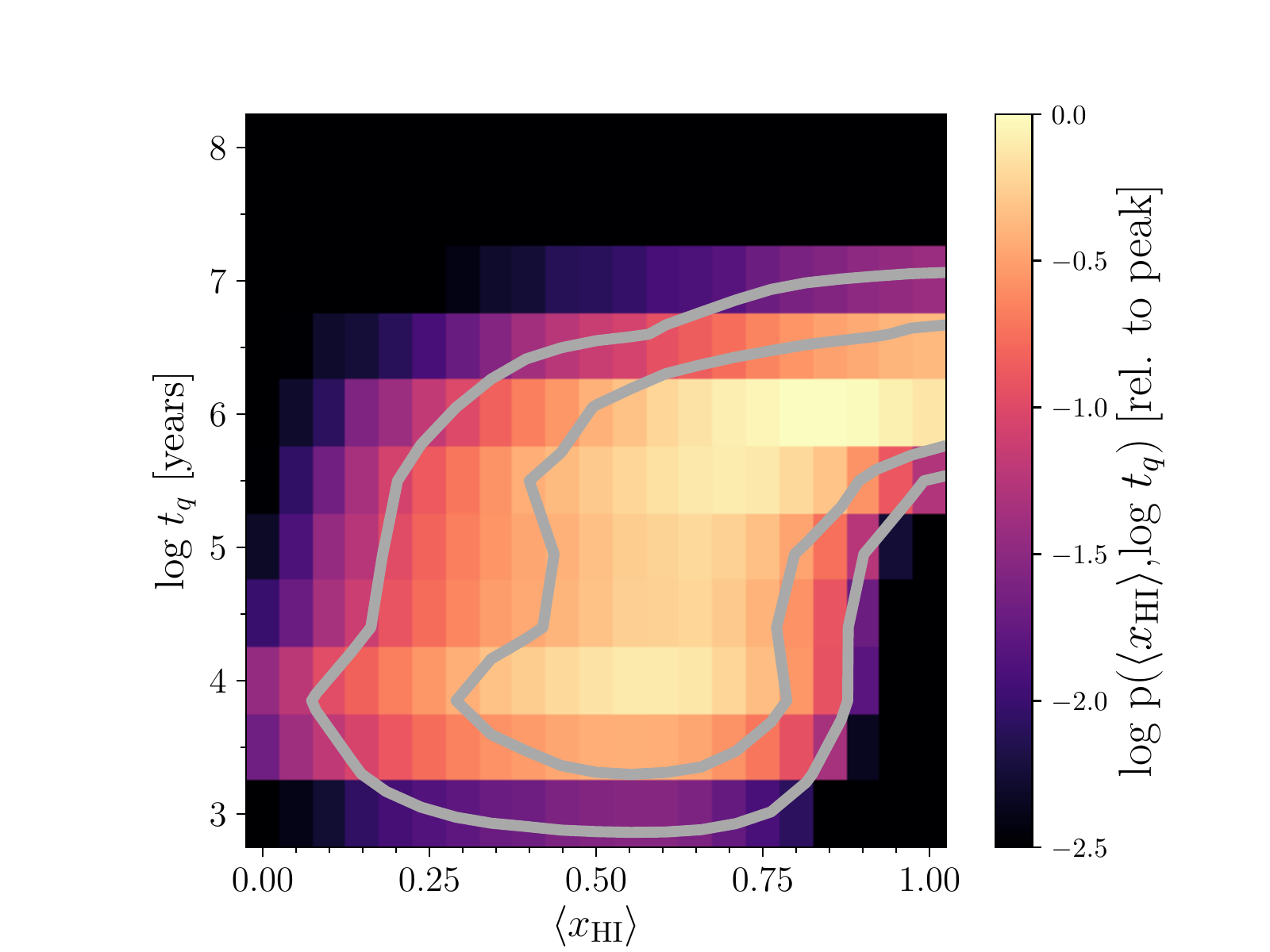}}\\
\end{center}
\caption{2D posterior PDF of $\langle x_{\rm HI}\rangle$ and $\log{t_{\rm q}}$ resulting from the M\~{L}E parameter values $\theta_{\rm M\tilde{L}E}=(\langle x_{\rm HI}\rangle=0.65,\log{t_{\rm q}}=6.5)$ derived from the ULAS J1342+0928 spectrum. The contours enclose $68\%$ and $95\%$ of the total probability.}
\label{fig:pisco_post}
\end{figure}

In Figure~\ref{fig:pisco_pca}, we show the red-side fit to the Magellan/FIRE + Gemini/GNIRS spectrum of ULAS J1342+0928 and the predicted blue-side (bias-corrected) continuum from Paper I. We find a best-fit template redshift of $z=7.4438$, a blueshift of ${\Delta}v=3422$ km/s from the systemic frame ($z=7.5413$, \citealt{Venemans17b}).
The red-side spectrum is very different from a typical quasar, however, similar examples do exist in our PCA training set (Paper I), and the PCA model is capable of broadly reproducing the spectrum. In fact, the uncertainty in the continuum derived from nearest-neighbor quasars in the training set is somewhat lower than for typical quasars due to the relatively weak broad emission lines.
The blue-side profile shows a strong damping wing redward of Ly$\alpha$, and a very small proximity zone ($R_p=1.20$ pMpc).
The damping wing signal is clearly stronger and the proximity zone is even smaller than ULAS J1120+0641, despite the slightly higher luminosity of ULAS J1342+0928. Both of these properties point towards a substantially neutral IGM surrounding ULAS J1342+0928.

We show the resulting transmission spectrum as the grey curve in Figure~\ref{fig:pisco_fit} and the 500 km/s-binned spectrum in black. No strong associated absorption is visible in the spectrum, so we include all pixels redward of Ly$\alpha$ that are covered by our modeled transmission spectra ($v-v_{\rm sys}<10,000$ km/s, $\lambda_{\rm rest}\la1255$ {\AA}).
We cut off the blue-side coverage at $+4000$ km/s ($\sim1200$ {\AA})
due to the presence of non-Gaussian noise features (both positive and negative spikes) in the spectrum that are not included in our noise model, which could spuriously impact the statistical analysis\footnote{This non-Gaussianity is likely also affecting the region of spectrum that we do analyze; however, we expect that even a modestly enhanced noise in the spectrum will still be subdominant compared to the other sources of variance that we consider (cosmic variance, continuum error).}. The maximum pseudo-likelihood parameter values are $\theta_{\rm M\tilde{L}{E}}= (\langle x_{\rm HI}\rangle=0.8, \log{t_{\rm q}}=6.0)$, and we compare the binned transmission spectrum to the median transmission profile of the M\~{L}E model (solid purple) and the expected 16--84th percentile scatter (dashed purple) from our forward modeling in Figure~\ref{fig:pisco_fit}.
From the M\~{L}E parameter values we infer the 2D posterior PDF $p(\theta|\theta_{\rm M\tilde{L}E})$
shown in Figure~\ref{fig:pisco_post}. Due in large part to the strong damping wing, the posterior PDF has a clear preference for a significantly neutral universe, even for short quasar lifetimes, although there is still some degeneracy between $\langle x_{\rm HI}\rangle$ and $t_{\rm q}$.

\section{Discussion}

In the preceding sections, we have demonstrated a method for jointly constraining the global neutral fraction $\langle x_{\rm HI}\rangle$ and quasar lifetime $t_{\rm q}$ from analysis of the proximity zone and (presence or absence)
damping wing, and applied it to the two quasars known at $z>7$: ULAS J1120+0641 at $z=7.09$, and ULAS J1342+0928 at $z=7.54$.
Here we present our constraints on $\langle x_{\rm HI}\rangle$ marginalized over quasar lifetime,
compare our analysis of ULAS J1120+0641 to
those from previous works, and discuss our choice of $\langle x_{\rm HI}\rangle$ and $t_{\rm q}$ priors.

\subsection{The History of Reionization: $\langle x_{\rm HI}\rangle(z)$}

\begin{figure}%[htb]
\begin{center}
\resizebox{8.50cm}{!}{\includegraphics[trim={1.5em 1em 4.0em 5.5em},clip]{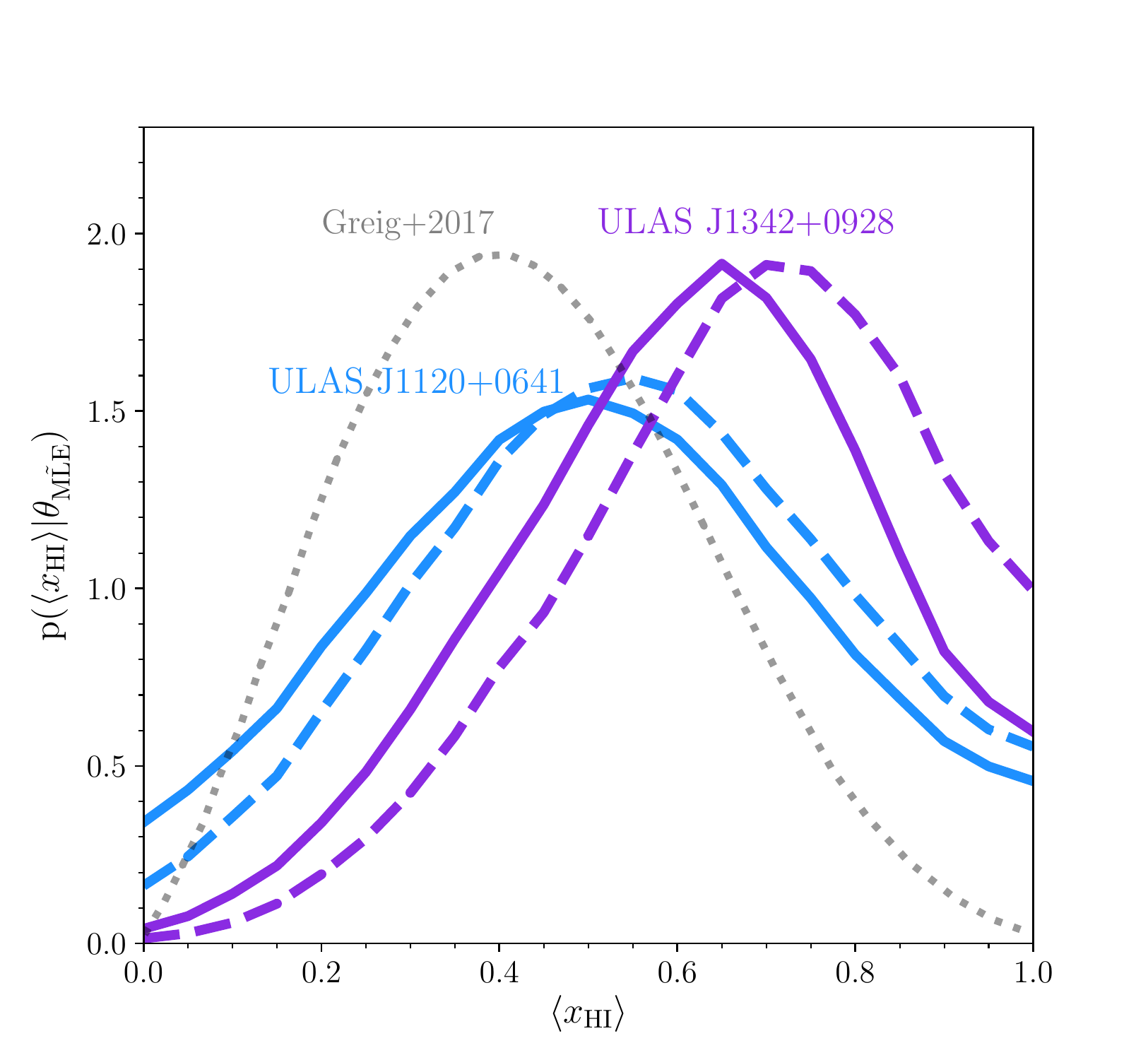}}\\
\end{center}
\caption{Posterior PDFs of $\langle x_{\rm HI}\rangle$ for ULAS J1120+0641 (blue) and ULAS J1342+0928 (purple) marginalized over quasar lifetime assuming a flat prior covering our entire model grid ($3.0 \leq \log{t_{\rm q}/{\rm yr}} \leq 8.0$; solid curves) or adopting a prior that excludes extremely short lifetimes ($5.0 \leq \log{t_{\rm q}/{\rm yr}} \leq 8.0$; dashed curves).
  The posterior PDF from the damping wing analysis of ULAS J1120+0641 in \citet{Greig17b} is shown as the dotted grey curve.}
\label{fig:xhi_pdf}
\end{figure}

In Figure~\ref{fig:xhi_pdf}, we show the posterior PDFs for $\langle x_{\rm HI}\rangle$ from each quasar, marginalized over quasar lifetime with a flat prior in log space from $10^3$ to $10^8$ years (solid curves) and a more restrictive prior from $10^5$ to $10^8$ years (dashed curves), and provide the 68\% and 95\% credible intervals in Table~\ref{tab:xhi}. We chose to include extremely short quasar lifetimes $\la10^5$ years in our fiducial analysis due to the existence of a surprisingly large fraction
of small proximity zones at $z\sim6$ ($\sim10\%$, \citealt{Eilers17}) which imply lifetimes shorter than $10^5$ years.
Including the possibility of the shortest quasar lifetimes shifts the $\langle x_{\rm HI}\rangle$ posterior PDF
to slightly lower values, consistent with the degeneracy between $\langle x_{\rm HI}\rangle$ and $t_{\rm q}$ shown in Figure~\ref{fig:model_stacks},
but in general does not have a large effect. While our analysis of ULAS J1120+0641 suggests a neutral fraction of $\langle x_{\rm HI} \rangle\sim0.5$, the posterior PDF is not particularly constraining, with significant probability density at the $\langle x_{\rm HI} \rangle=0$ and $\langle x_{\rm HI} \rangle=1$ boundaries. In contrast, the posterior PDF for ULAS J1342+0928 strongly indicates a significantly neutral IGM, ruling out $\langle x_{\rm HI} \rangle < 0.08\ (0.14)$ at 99\% probability marginalized over quasar lifetime for our prior covering $10^3 \leq t_{\rm q} \leq 10^8$ ($10^5 \leq t_{\rm q} \leq 10^8$) years.

\begin{figure}%[htb]
\begin{center}
\resizebox{8.50cm}{!}{\includegraphics[trim={1.0em 1em 4.0em 5.0em},clip]{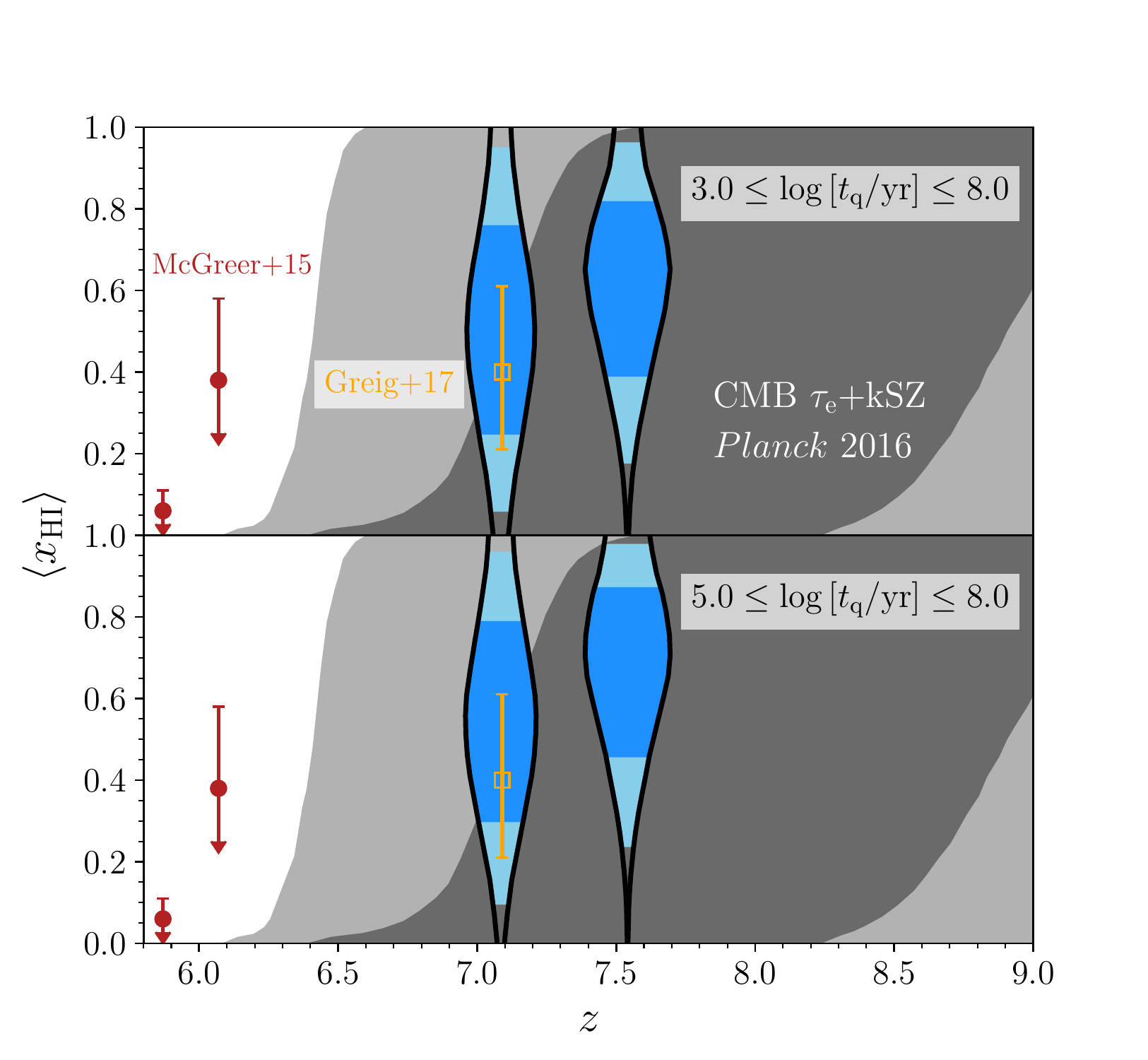}}
\end{center}
\caption{Violin plot comparing the posterior PDFs from our analysis with the reionization history constraints from \citet{Planck16b}, with the dark and light grey shaded regions corresponding to the 68\% and 95\% credible intervals, respectively. Also shown are the Ly$\alpha$+Ly$\beta$ forest dark pixel constraints from \citet{McGreer15} (red crosses) and the damping wing analysis of ULAS J1120+0641 from \citet{Greig17b} (orange square).}
\label{fig:planck}
\end{figure}

In Figure~\ref{fig:planck}, we compare the posterior PDFs for $\langle x_{\rm HI}\rangle$ from each quasar to the broad swathe of reionization histories consistent with the measured electron-scattering optical depth of the
CMB \citep{Planck16b}\footnote{We compare to the combined Planck + ACT/SPT constraints on the reionization history that take into account upper limits on the strength of the kinetic Sunyaev-Zel'dovich effect and a prior from Ly$\alpha$ forest measurements that the end of reionization occurred before $z=6$, see \S~5.3 of \citet{Planck16b}}.
Under our most conservative prior, allowing quasar lifetimes from $10^3$ to $10^8$ years, we find 68\% (95\%) credible intervals of $\langle x_{\rm HI}\rangle(z=7.09)=0.48^{+0.26}_{-0.26}(^{+0.47}_{-0.46})$ and $\langle x_{\rm HI}\rangle(z=7.54)=0.60^{+0.20}_{-0.23}(^{+0.36}_{-0.45})$. These constraints are consistent with the CMB and are in broad agreement with recent calculations of the reionization history (e.g. \citealt{Robertson15,Bouwens15,Khaire16}). %, although this may not be particularly surprising given our large uncertainties.
The large cosmic variance between the damping wing profiles at fixed $\langle x_{\rm HI}\rangle$ (\S~\ref{sec:model}), the strong degeneracy with quasar lifetime,
and the limited precision of our continuum reconstructions
greatly limits the constraining power of any single $z>7$ quasar. However, with only a handful of additional quasars at $z>7$, it may be possible to constrain $\langle x_{\rm HI} \rangle(z)$
to $\sim10\%$. That said, despite the substantial uncertainties, our analysis of two $z>7$ quasars already constrains the reionization history more than the integral constraint from the CMB.

\begin{table*}[tb]
\begin{center}
\caption{Neutral fraction constraints from the proximity zone and damping wing}
\label{tab:xhi}
\begin{tabular}{c c c c c}
\hline \noalign {\smallskip}
Quasar & $z$ & $\langle x_{\rm HI}\rangle$ $(10^3 \leq t_{\rm q}/{\rm yr} \leq 10^8)$ & $\langle x_{\rm HI}\rangle$ $(10^5 \leq t_{\rm q}/{\rm yr} \leq 10^8)$ \\
\hline \noalign {\smallskip}
ULAS J1120+0641 & 7.0851 & $0.48^{+0.26}_{-0.26}(^{+0.47}_{-0.46})$ & $0.52^{+0.25}_{-0.25}(^{+0.44}_{-0.46})$ \\
ULAS J1342+0928 & 7.5413 & $0.60^{+0.20}_{-0.23}(^{+0.36}_{-0.45})$ & $0.67^{+0.19}_{-0.23}(^{+0.31}_{-0.45})$ \\
\hline \noalign {\smallskip}
\end{tabular}
\end{center}
The tabulated constraints represent the median and 68\% (95\%) credible intervals obtained via linear interpolation of the $t_{\rm q}$-marginalized posterior PDFs in Figure~\ref{fig:xhi_pdf}.
\end{table*}

\subsection{Previous studies of ULAS J1120+0641}

\begin{figure}%[htb]
\begin{center}
\resizebox{8.50cm}{!}{\includegraphics[trim={1.0em 0em 3.5em 3em},clip]{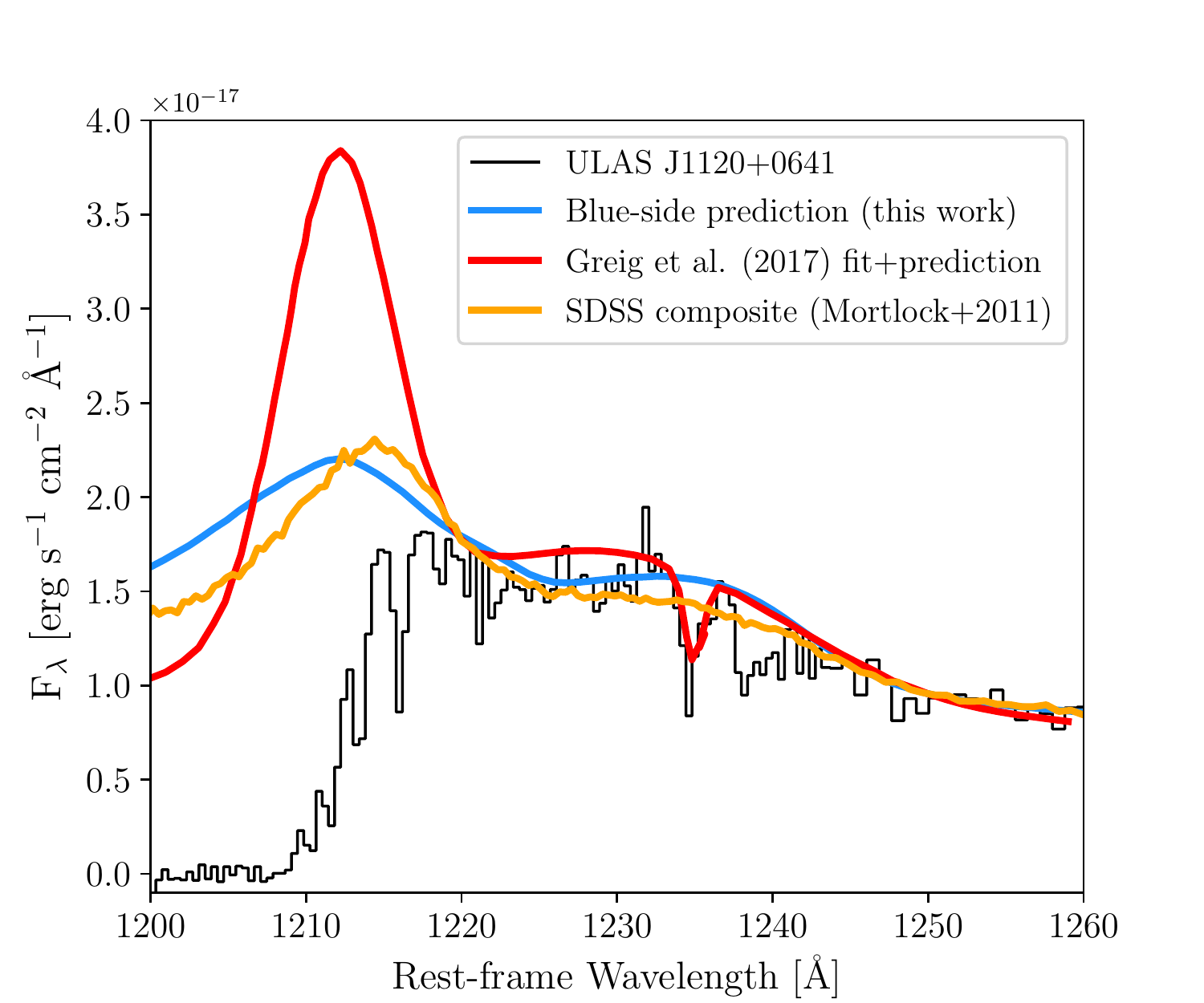}}
\end{center}
\caption{Comparison between our blue-side prediction for ULAS J1120+0641 (blue), the SDSS matched composite spectrum from \citet{Mortlock11} (orange), and the model from \citet{Greig17b}. The \citet{Greig17b} model has been renormalized to our blue-side prediction at $\lambda_{\rm rest}=1245$ {\AA} to correct for the different flux calibration of the \citet{Simcoe12} spectrum used in their analysis.}
\label{fig:greig}
\end{figure}

In the original discovery paper for ULAS J1120+0641, \citet{Mortlock11} suggested that the spectrum showed signs of an IGM damping wing. They selected a sample of lower-redshift quasars with similar \ion{C}{4} blueshifts (relative to \ion{Mg}{2}) and equivalent widths, and stacked their spectra to predict the shape of ULAS J1120+0641. The resulting composite spectrum was somewhat above the observed spectrum at wavelengths at and just redward of rest-frame Ly$\alpha$, with a shape resembling the characteristic damping wing profile. However, the uncertainty in the stacked composite was not fully quantified, and the physical model was limited to the \citet{ME98} expression for the damping wing. A followup work by \citet{Bolton11} expanded upon the physical model with 1D radiative transfer simulations, and found that the combination of absorption at rest-frame Ly$\alpha$ and the small proximity zone were suggestive of neutral gas close to the quasar ($x_{\rm HI}\sim0.1$), although they noted that an identical signal could potentially come from small-scale optically thick gas along the line of sight instead of a neutral IGM (see also \citealt{Keating15}). With a higher resolution FIRE spectrum, \citet{Simcoe12} found that any such dense gas would have to be extremely metal-poor ($[Z/H]<-4$), which would seem to favor the IGM interpretation.

The accuracy of the \citet{Mortlock11} composite spectrum as a prediction for ULAS J1120+0641 was called into question by \citet{BB15}, because the composite spectrum fails to match the \ion{C}{4} line and this may lead to an overestimate of the Ly$\alpha$ emission. \citet{BB15} selected a comparison sample of low-redshift quasars with more precisely-matched \ion{C}{4} emission line profiles. They found that the shape of the Ly$\alpha$+\ion{N}{5} region of these spectra was nearly identical to ULAS J1120+0641, suggesting that there may not be any damped absorption at all.

The most recent analysis of the ULAS J1120+0641 damping wing profile was undertaken by \citet{Greig17b}. Similar to this work, they trained a predictive model for the intrinsic blue-side continuum from a large sample of BOSS quasar spectra \citep{Greig17a}. Their parametric model predicts Gaussian emission line parameters for Ly$\alpha$ (line width, amplitude, and velocity shift of two components) from fits to several broad emission lines on the red side of the spectrum. In addition, they separately fit the \ion{N}{5}+\ion{Si}{2} complex at $1230 < \lambda_{\rm rest} < 1275$ {\AA} and introduce this fit as a strict prior to the model for the Ly$\alpha$ damping wing region ($1218 < \lambda_{\rm rest} < 1230$ {\AA}). Again similar to our analysis, \citet{Greig17b} employed a large-volume semi-numerical simulation of reionization (the Evolution Of 21 cm Structure simulation, \citealt{Mesinger16}) to characterize the large-scale distribution of neutral gas around massive halos during the reionization epoch. By restricting their analysis to wavelengths redward of Ly$\alpha$, they did not need to explicitly model the proximity zone of the quasar.
To approximate the effect of the quasar ionizing radiation on neutral gas along the line of sight, they ionize the first 16 comoving Mpc ($\sim2$ proper Mpc) of every sightline to be consistent with the observed profile,
but beyond this distance the quasar does not affect the ionization topology.
Their final statistical constraints on $\langle x_{\rm HI}\rangle$ were derived from a $\chi^2$-based likelihood analysis of intrinsic Ly$\alpha$ profiles drawn from their predictive
continuum model multiplied by the damping wing absorption from sightlines through their reionization simulation. While \citet{Greig17b} presented the most sophisticated study of a quasar damping wing at the time, their method has a handful of potential shortcomings which we describe below.

First, their fit of the \ion{N}{5} and \ion{Si}{2} complex and subsequent prior on the spectrum begins at $\lambda_{\rm rest}=1230$ {\AA} under the assumption that the damping wing absorption there is minimal. However, the absorption at $\lambda_{\rm rest}\sim1230$ {\AA} can still be significant at large $\langle x_{\rm HI}\rangle$, a fact which can be readily seen in the inset panel of Figure 2 in \citet{Greig17b} where their best-fit damping wing model still shows $>2\sigma$
absorption at $\lambda_{\rm rest}\sim1230$ {\AA} (see also Figure~\ref{fig:model_stacks}).
This prior may then result in a bias towards lower $\langle x_{\rm HI}\rangle$, as smooth absorption redward of $\lambda_{\rm rest}\sim1230$ {\AA} may be fitted out, although it is unclear what the magnitude of this effect would be in practice. 

Additionally, while they account for scatter in the Gaussian parameters for the Ly$\alpha$ line, they do not appear to account for any error in the Gaussian fits themselves. Indeed, as described in \citet{Greig17a}, they remove quasars from the training set whose spectra are ``not well fit or characterized" by their double-Gaussian model for the Ly$\alpha$ line. While we have also excluded some discrepant quasars from our analysis, they were exclusively
the most extreme cases of BALs and associated absorption which would be readily apparent in the spectra of $z>7$ quasars. The distribution of possible continua for ULAS J1120+0641 used in \citet{Greig17b} thus represents an underestimate of the true continuum error, lacking the additional error resulting from any deviations in the true continuum from the multiple Gaussian model.

Finally, by always treating the first $\sim2$ proper Mpc of every sightline as ionized, \citet{Greig17b} introduce a complicated prior on the quasar lifetime. For sightlines which originally intersect neutral gas within 2 proper Mpc, the quasar must have been on long enough to ionize material out to that distance. However, for sightlines where the first 2 proper Mpc are already ionized, the quasar then has no effect at all on neutral gas along the line of sight, implying a very short lifetime. As such the \citet{Greig17b} posterior PDF cannot be considered fully marginalized over quasar lifetime, making a direct comparison between our $t_{\rm q}$-marginalized $\langle x_{\rm HI}\rangle$ constraints (Figure~\ref{fig:xhi_pdf}) and the results of the \citet{Greig17b} analysis very difficult.

In Figure~\ref{fig:greig}, we compare our blue-side prediction for ULAS J1120+0641 to the \citet{Mortlock11} composite spectrum and the \citet{Greig17b} model, where the latter has been renormalized to match our prediction at $\lambda_{\rm rest}=1245$ {\AA}. While our method predicts a very similar continuum to \citet{Mortlock11}, the intrinsic Ly$\alpha$ emission line strength predicted by \citet{Greig17b} is dramatically higher (albeit with nearly identical Ly$\alpha$ centroids). Most importantly, however, we predict a substantially lower continuum (i.e. much closer to the observed spectrum) at $\lambda_{\rm rest}\sim1225$ {\AA}, the spectral region which contributed the most to the damping wing detection in \citet{Greig17b}. As a result, our measurement does not rule out $\langle x_{\rm HI}\rangle\sim0$, although we nevertheless prefer somewhat higher $\langle x_{\rm HI}\rangle$ than \citet{Greig17b}
(see Figure~\ref{fig:xhi_pdf}), in large part due to the small proximity zone. While it is currently unclear why our two methods predict substantially different continua, we note that our predicted continuum is more consistent with the weak Ly$\alpha$ lines of the ULAS J1120+0641-analogs discussed in \citet{BB15}, and with the nearest-neighbor quasars we used to calibrate the continuum uncertainty in Paper I.

\subsection{Choice of $\langle x_{\rm HI}\rangle$ and $t_q$ Priors}\label{sec:priors}

As mentioned above, we assume a flat prior on $\langle x_{\rm HI}\rangle$ from ``0" (in truth, a model where reionization has finished with residual $\langle x_{\rm HI}\rangle\sim10^{-3}$) to 1.0. If we were to instead assume a flat logarithmic prior (i.e. $p(\langle x_{\rm HI} \rangle) \propto 1/\langle x_{\rm HI} \rangle$) that extended down to $\langle x_{\rm HI}\rangle\sim10^{-4}$, which would still be consistent with a completely opaque Ly$\alpha$ forest at
$z\ga7$,
our constraints on $\langle x_{\rm HI}\rangle$ would be dragged down to $\langle x_{\rm HI}\rangle\sim0$, and there would be little evidence at all for ongoing reionization -- the posterior PDF at $\langle x_{\rm HI}\rangle\sim0$, which is small but non-negligible (Figure~\ref{fig:xhi_pdf}), would be boosted by a factor of $>10^3$ relative to $\langle x_{\rm HI}\rangle>0.1$. Because the damping wing signal is only detected at modest
statistical significance in the context of the covariant continuum errors in our
PCA method (even for ULAS J1342+0928), switching to a
prior on $\langle x_{\rm HI}\rangle$ that is instead uniform in log space would shift the posterior PDF to peak close to $\langle x_{\rm HI}\rangle\sim0$ for essentially \emph{any} realistic quasar damping wing signal unless $\langle x_{\rm HI}\rangle\sim1$ and $t_{\rm q}$ is very short.
One could argue that our prior knowledge of $\langle x_{\rm HI} \rangle$ is not simply log-uniform, however, but can instead be thought of as bimodal.
If reionization is complete, the Universe is ``highly ionized," with $\langle x_{\rm HI}\rangle\la10^{-3}$ set by photoionization equilibrium with a metagalactic ionizing radiation field (e.g. \citealt{HM12}). If the Universe is undergoing the reionization phase transition, we instead have $\langle x_{\rm HI} \rangle$ of order unity.

While we are still starved for $z>7$ quasars we will remain in a regime where quantitative constraints on $\langle x_{\rm HI} \rangle$ depend strongly on our choice of priors, but larger samples will greatly reduce this dependence: the prior enters the posterior PDF only once while each additional quasar contributes to the likelihood function. Assuming a Gaussian likelihood with $\langle x_{\rm HI}\rangle=0.5$ and $\sigma_{\langle x_{\rm HI}\rangle}\sim0.3$, similar to our constraint from the ULAS J1120+0641 spectrum, to counteract a factor of $10^4$ prior advantage at $\langle x_{\rm HI}\rangle\sim10^{-4}$ would require $\ga7$ quasars. Such a population of $z\ga7$ quasars is within reach of current programs exploiting wide-field optical and near-infrared surveys (e.g. \citealt{Wang17,Banados18}). 

Our fiducial prior on $t_q$ is a flat, logarithmic prior from $10^3$ to $10^8$ years. The lower limit is motivated by the extremely small proximity zones in \citet{Eilers17} whose sizes are consistent with such a short lifetime ($t_{\rm q}<10^5$ years).
The upper limit comes from the fact that, assuming accretion at the Eddington limit with $10\%$ radiative efficiency, $10^8$ years ago the quasar would have been roughly an order of magnitude fainter than currently observed. The quasar would then be effectively shut off at early times, so longer lifetimes would be largely irrelevant to the proximity zone and damping wing structure and can thus be excluded. Other estimates from the thermal proximity effect \citep{Bolton12} and \ion{He}{2} transverse proximity effect \citep{Schmidt17a,Schmidt17b} suggest lifetimes of at least $10^7$ years.
The spectra of ULAS J1120+0641 and ULAS J1342+0928 both appear to exclude lifetimes at the upper and lower ends of our fiducial prior range (Figures \ref{fig:mortlock_post} and \ref{fig:pisco_post}), however, so expanding the bounds in either direction would make little difference to our posterior PDFs.

The choice of flat prior in linear space vs. log space for $t_{\rm q}$ is more subtle.
If one assumes that all quasars live for a fixed amount of time $t_{\rm q,max}$, then a random quasar will have been shining continuously for a time $t_{\rm q}$
(which is what we have defined as ``lifetime" in this work) drawn from a uniform distribution between $0 < t_{\rm q} < t_{\rm q,max}$, and so a flat prior in linear space would be appropriate. However, the uncertainty on this maximum lifetime spans multiple orders of magnitude (e.g. \citealt{Martini04}), so we believe that a flat prior in log space is reasonably well justified. 

\section{Conclusion}

In this work we have used the intrinsic quasar continuum models of ULAS J1120+0641 and ULAS J1342+0928 from Paper I, in combination with extensive forward modeling of the proximity zone and damping wing features in the context of patchy reionization, to jointly constrain the lifetimes of the two quasars and the volume-averaged neutral fraction of the Universe at $z>7$.

Our hybrid model of quasar spectra combines large-scale semi-numerical reionization simulations, hydrodynamical simulations, and 1D radiative transfer of ionizing photons from the quasars. We computed 2400 transmission spectra covering the proximity zone and damping wing for each pair of $\langle x_{\rm HI}\rangle$ and $\log{t_{\rm q}}$ on a coarse $21\times11$ grid for both quasars. Accounting for the covariant intrinsic quasar continuum uncertainty from Paper I, we can then construct realistic forward modeled representations of quasar transmission spectra. Based
on these mock spectra we developed a Bayesian statistical method for recovering the joint posterior PDF of $\langle x_{\rm HI}\rangle$ and $\log{t_{\rm q}}$ from an observed quasar spectrum.

Applying our statistical methodology to the spectra of ULAS J1120+0641 at $z=7.09$ \citep{Mortlock11} and ULAS J1342+0928 at $z=7.54$ \citep{Banados18}, we found that both quasars are consistent with an ongoing epoch of reionization at $z>7$. When marginalized over quasar lifetimes from $10^3$ to $10^8$ years, the resulting medians and 68\% credible intervals of the posterior PDFs are $\langle x_{\rm HI}\rangle(z=7.09)=0.48^{+0.26}_{-0.26}$ and $\langle x_{\rm HI}\rangle(z=7.54)=0.60^{+0.20}_{-0.23}$.

Using our method it should be possible to constrain $\langle x_{\rm HI}\rangle$ at lower redshifts $z\sim6$--$7$ where there are far more quasars known (e.g. \citealt{Venemans13,Venemans15,Banados16,Reed17,Wang17,Mazzucchelli17}). The most constraining measurements at lower redshift to date are the model-independent upper limits from \citet{McGreer15} who measured the fraction of dark pixels in the co-spatial Ly$\alpha$ and Ly$\beta$ forests, shown as red points in Figure~\ref{fig:planck}, but this method becomes less constraining as the Ly-series forests become almost entirely opaque at $z\ga6$, which may simply result from density evolution in the IGM or a mild decrease in the ionizing background towards higher redshift (e.g. \citealt{Davies17}).
We predict that we will be able to obtain stronger constraints than the $z\sim6.1$ upper limit of \citet{McGreer15} from the multitude of existing spectra of quasars at $z\sim6-7$, an endeavor we leave for future work.

Large samples of $z\ga7$ quasars to be discovered in further follow-up of quasar candidates from ground-based surveys (e.g. ULAS, \citealt{Lawrence07}; VIKING, \citealt{Arnaboldi07}; VHS, \citealt{McMahon13}; DECaLS\footnote{\url{http://legacysurvey.org/}}; UHS, \citealt{Dye18}) and in future wide-field near-infrared surveys by Euclid and WFIRST, together with high signal-to-noise spectra from JWST, will allow for exquisitely precise constraints on $\langle x_{\rm HI}\rangle(z)$.

\section*{Acknowledgements}
We would like to thank A. Price-Whelan and D. Hogg for consultation on statistical methods, and D. Stern for supporting the discovery of ULAS J1342+0928.

BPV and F. Walter acknowledge funding through the ERC grants ``Cosmic Dawn" and ``Cosmic Gas".

 \newcommand{\noop}[1]{}


\begin{thebibliography}{}
\expandafter\ifx\csname natexlab\endcsname\relax\def\natexlab#1{#1}\fi

\bibitem[{{Almgren} {et~al.}(2013){Almgren}, {Bell}, {Lijewski}, {Luki{\'c}},
  \& {Van Andel}}]{Almgren13}
{Almgren}, A.~S., {Bell}, J.~B., {Lijewski}, M.~J., {Luki{\'c}}, Z., \& {Van
  Andel}, E. 2013, \apj, 765, 39

\bibitem[{{Alvarez} \& {Abel}(2007)}]{AA07}
{Alvarez}, M.~A., \& {Abel}, T. 2007, \mnras, 380, L30

\bibitem[{{Arnaboldi} {et~al.}(2007){Arnaboldi}, {Neeser}, {Parker}, {Rosati},
  {Lombardi}, {Dietrich}, \& {Hummel}}]{Arnaboldi07}
{Arnaboldi}, M., {Neeser}, M.~J., {Parker}, L.~C., {et~al.} 2007, The
  Messenger, 127

\bibitem[{{Ba{\~n}ados} {et~al.}(2016){Ba{\~n}ados}, {Venemans}, {Decarli},
  {Farina}, {Mazzucchelli}, {Walter}, {Fan}, {Stern}, {Schlafly}, {Chambers},
  {Rix}, {Jiang}, {McGreer}, {Simcoe}, {Wang}, {Yang}, {Morganson}, {De Rosa},
  {Greiner}, {Balokovi{\'c}}, {Burgett}, {Cooper}, {Draper}, {Flewelling},
  {Hodapp}, {Jun}, {Kaiser}, {Kudritzki}, {Magnier}, {Metcalfe}, {Miller},
  {Schindler}, {Tonry}, {Wainscoat}, {Waters}, \& {Yang}}]{Banados16}
{Ba{\~n}ados}, E., {Venemans}, B.~P., {Decarli}, R., {et~al.} 2016, \apjs, 227,
  11

\bibitem[{{Ba{\~n}ados} {et~al.}(2018){Ba{\~n}ados}, {Venemans},
  {Mazzucchelli}, {Farina}, {Walter}, {Wang}, {Decarli}, {Stern}, {Fan},
  {Davies}, {Hennawi}, {Simcoe}, {Turner}, {Rix}, {Yang}, {Kelson}, {Rudie}, \&
  {Winters}}]{Banados18}
{Ba{\~n}ados}, E., {Venemans}, B.~P., {Mazzucchelli}, C., {et~al.} 2018, \nat,
  553, 473

\bibitem[{{Becker} {et~al.}(2001){Becker}, {Fan}, {White}, {Strauss},
  {Narayanan}, {Lupton}, {Gunn}, {Annis}, {Bahcall}, {Brinkmann}, {Connolly},
  {Csabai}, {Czarapata}, {Doi}, {Heckman}, {Hennessy}, {Ivezi{\'c}}, {Knapp},
  {Lamb}, {McKay}, {Munn}, {Nash}, {Nichol}, {Pier}, {Richards}, {Schneider},
  {Stoughton}, {Szalay}, {Thakar}, \& {York}}]{Becker01}
{Becker}, R.~H., {Fan}, X., {White}, R.~L., {et~al.} 2001, \aj, 122, 2850

\bibitem[{{Bolton} {et~al.}(2012){Bolton}, {Becker}, {Raskutti}, {Wyithe},
  {Haehnelt}, \& {Sargent}}]{Bolton12}
{Bolton}, J.~S., {Becker}, G.~D., {Raskutti}, S., {et~al.} 2012, \mnras, 419,
  2880

\bibitem[{{Bolton} \& {Haehnelt}(2007)}]{BH07}
{Bolton}, J.~S., \& {Haehnelt}, M.~G. 2007, \mnras, 374, 493

\bibitem[{{Bolton} {et~al.}(2011){Bolton}, {Haehnelt}, {Warren}, {Hewett},
  {Mortlock}, {Venemans}, {McMahon}, \& {Simpson}}]{Bolton11}
{Bolton}, J.~S., {Haehnelt}, M.~G., {Warren}, S.~J., {et~al.} 2011, \mnras,
  416, L70

\bibitem[{{Bosman} \& {Becker}(2015)}]{BB15}
{Bosman}, S.~E.~I., \& {Becker}, G.~D. 2015, \mnras, 452, 1105

\bibitem[{{Bouwens} {et~al.}(2015){Bouwens}, {Illingworth}, {Oesch}, {Trenti},
  {Labb{\'e}}, {Bradley}, {Carollo}, {van Dokkum}, {Gonzalez}, {Holwerda},
  {Franx}, {Spitler}, {Smit}, \& {Magee}}]{Bouwens15}
{Bouwens}, R.~J., {Illingworth}, G.~D., {Oesch}, P.~A., {et~al.} 2015, \apj,
  803, 34

\bibitem[{{Carswell} {et~al.}(1982){Carswell}, {Whelan}, {Smith}, {Boksenberg},
  \& {Tytler}}]{Carswell82}
{Carswell}, R.~F., {Whelan}, J.~A.~J., {Smith}, M.~G., {Boksenberg}, A., \&
  {Tytler}, D. 1982, \mnras, 198, 91

\bibitem[{{Cen} \& {Haiman}(2000)}]{CH00}
{Cen}, R., \& {Haiman}, Z. 2000, \apjl, 542, L75

\bibitem[{{Dall'Aglio} {et~al.}(2008){Dall'Aglio}, {Wisotzki}, \&
  {Worseck}}]{Dall'Aglio08}
{Dall'Aglio}, A., {Wisotzki}, L., \& {Worseck}, G. 2008, \aap, 491, 465

\bibitem[{{Davies} {et~al.}(2016){Davies}, {Furlanetto}, \&
  {McQuinn}}]{Davies16}
{Davies}, F.~B., {Furlanetto}, S.~R., \& {McQuinn}, M. 2016, \mnras, 457, 3006

\bibitem[{{Davies} {et~al.}(2017){Davies}, {Hennawi}, {Eilers}, \&
  {Luki{\'c}}}]{Davies17}
{Davies}, F.~B., {Hennawi}, J.~F., {Eilers}, A.-C., \& {Luki{\'c}}, Z. 2017,
  ArXiv e-prints, arXiv:1703.10174

\bibitem[{{Davies} {et~al.}(2018){Davies}, {Hennawi}, {Ba{\~n}ados}, {Simcoe},
  {Decarli}, {Fan}, {Farina}, {Mazzucchelli}, {Rix}, {Venemans}, {Walter},
  {Wang}, \& {Yang}}]{Davies18a}
{Davies}, F.~B., {Hennawi}, J.~F., {Ba{\~n}ados}, E., {et~al.} 2018, ArXiv
  e-prints, arXiv:1801.07679

\bibitem[{Drovandi {et~al.}(2015)Drovandi, Pettitt, \& Lee}]{Drovandi15}
Drovandi, C.~C., Pettitt, A.~N., \& Lee, A. 2015, Statist. Sci., 30, 72

\bibitem[{{Dye} {et~al.}(2018){Dye}, {Lawrence}, {Read}, {Fan}, {Kerr},
  {Varricatt}, {Furnell}, {Edge}, {Irwin}, {Hambly}, {Lucas}, {Almaini},
  {Chambers}, {Green}, {Hewett}, {Liu}, {McGreer}, {Best}, {Zhang}, {Sutorius},
  {Froebrich}, {Magnier}, {Hasinger}, {Lederer}, {Bold}, \& {Tedds}}]{Dye18}
{Dye}, S., {Lawrence}, A., {Read}, M.~A., {et~al.} 2018, \mnras, 473, 5113

\bibitem[{{Eilers} {et~al.}(2017){Eilers}, {Davies}, {Hennawi}, {Prochaska},
  {Luki{\'c}}, \& {Mazzucchelli}}]{Eilers17}
{Eilers}, A.-C., {Davies}, F.~B., {Hennawi}, J.~F., {et~al.} 2017, \apj, 840,
  24

\bibitem[{{Fan} {et~al.}(2001){Fan}, {Narayanan}, {Lupton}, {Strauss}, {Knapp},
  {Becker}, {White}, {Pentericci}, {Leggett}, {Haiman}, {Gunn}, {Ivezi{\'c}},
  {Schneider}, {Anderson}, {Brinkmann}, {Bahcall}, {Connolly}, {Csabai}, {Doi},
  {Fukugita}, {Geballe}, {Grebel}, {Harbeck}, {Hennessy}, {Lamb}, {Miknaitis},
  {Munn}, {Nichol}, {Okamura}, {Pier}, {Prada}, {Richards}, {Szalay}, \&
  {York}}]{Fan01}
{Fan}, X., {Narayanan}, V.~K., {Lupton}, R.~H., {et~al.} 2001, \aj, 122, 2833

\bibitem[{{Fan} {et~al.}(2003){Fan}, {Strauss}, {Schneider}, {Becker}, {White},
  {Haiman}, {Gregg}, {Pentericci}, {Grebel}, {Narayanan}, {Loh}, {Richards},
  {Gunn}, {Lupton}, {Knapp}, {Ivezi{\'c}}, {Brandt}, {Collinge}, {Hao},
  {Harbeck}, {Prada}, {Schaye}, {Strateva}, {Zakamska}, {Anderson},
  {Brinkmann}, {Bahcall}, {Lamb}, {Okamura}, {Szalay}, \& {York}}]{Fan03}
{Fan}, X., {Strauss}, M.~A., {Schneider}, D.~P., {et~al.} 2003, \aj, 125, 1649

\bibitem[{{Fan} {et~al.}(2006){Fan}, {Strauss}, {Becker}, {White}, {Gunn},
  {Knapp}, {Richards}, {Schneider}, {Brinkmann}, \& {Fukugita}}]{Fan06}
{Fan}, X., {Strauss}, M.~A., {Becker}, R.~H., {et~al.} 2006, \aj, 132, 117

\bibitem[{{Furlanetto} \& {Johnson Stoever}(2010)}]{FJS10}
{Furlanetto}, S.~R., \& {Johnson Stoever}, S. 2010, \mnras, 404, 1869

\bibitem[{{Furlanetto} {et~al.}(2006){Furlanetto}, {Oh}, \&
  {Briggs}}]{Furlanetto06}
{Furlanetto}, S.~R., {Oh}, S.~P., \& {Briggs}, F.~H. 2006, \physrep, 433, 181

\bibitem[{{Furlanetto} {et~al.}(2004){Furlanetto}, {Zaldarriaga}, \&
  {Hernquist}}]{Furlanetto04}
{Furlanetto}, S.~R., {Zaldarriaga}, M., \& {Hernquist}, L. 2004, \apj, 613, 1

\bibitem[{{Gnedin} \& {Hui}(1998)}]{GH98}
{Gnedin}, N.~Y., \& {Hui}, L. 1998, \mnras, 296, 44

\bibitem[{Gourieroux {et~al.}(1993)Gourieroux, Monfort, \&
  Renault}]{Gourieroux93}
Gourieroux, C., Monfort, A., \& Renault, E. 1993, Journal of Applied
  Econometrics, 8, S85

\bibitem[{{Greig} {et~al.}(2017{\natexlab{a}}){Greig}, {Mesinger}, {Haiman}, \&
  {Simcoe}}]{Greig17b}
{Greig}, B., {Mesinger}, A., {Haiman}, Z., \& {Simcoe}, R.~A.
  2017{\natexlab{a}}, \mnras, 466, 4239

\bibitem[{{Greig} {et~al.}(2017{\natexlab{b}}){Greig}, {Mesinger}, {McGreer},
  {Gallerani}, \& {Haiman}}]{Greig17a}
{Greig}, B., {Mesinger}, A., {McGreer}, I.~D., {Gallerani}, S., \& {Haiman}, Z.
  2017{\natexlab{b}}, \mnras, 466, 1814

\bibitem[{{Gunn} \& {Peterson}(1965)}]{GP65}
{Gunn}, J.~E., \& {Peterson}, B.~A. 1965, \apj, 142, 1633

\bibitem[{{Haardt} \& {Madau}(2012)}]{HM12}
{Haardt}, F., \& {Madau}, P. 2012, \apj, 746, 125

\bibitem[{{Keating} {et~al.}(2015){Keating}, {Haehnelt}, {Cantalupo}, \&
  {Puchwein}}]{Keating15}
{Keating}, L.~C., {Haehnelt}, M.~G., {Cantalupo}, S., \& {Puchwein}, E. 2015,
  \mnras, 454, 681

\bibitem[{{Khaire} {et~al.}(2016){Khaire}, {Srianand}, {Choudhury}, \&
  {Gaikwad}}]{Khaire16}
{Khaire}, V., {Srianand}, R., {Choudhury}, T.~R., \& {Gaikwad}, P. 2016,
  \mnras, 457, 4051

\bibitem[{{Khrykin} {et~al.}(2017){Khrykin}, {Hennawi}, \&
  {McQuinn}}]{Khrykin17}
{Khrykin}, I.~S., {Hennawi}, J.~F., \& {McQuinn}, M. 2017, \apj, 838, 96

\bibitem[{{Khrykin} {et~al.}(2016){Khrykin}, {Hennawi}, {McQuinn}, \&
  {Worseck}}]{Khrykin16}
{Khrykin}, I.~S., {Hennawi}, J.~F., {McQuinn}, M., \& {Worseck}, G. 2016, \apj,
  824, 133

\bibitem[{{Kramer} \& {Haiman}(2009)}]{KH09}
{Kramer}, R.~H., \& {Haiman}, Z. 2009, \mnras, 400, 1493

\bibitem[{{Kulkarni} {et~al.}(2015){Kulkarni}, {Hennawi}, {O{\~n}orbe},
  {Rorai}, \& {Springel}}]{Kulkarni15}
{Kulkarni}, G., {Hennawi}, J.~F., {O{\~n}orbe}, J., {Rorai}, A., \& {Springel},
  V. 2015, \apj, 812, 30

\bibitem[{{Lacey} \& {Cole}(1993)}]{LC93}
{Lacey}, C., \& {Cole}, S. 1993, \mnras, 262, 627

\bibitem[{{Lawrence} {et~al.}(2007){Lawrence}, {Warren}, {Almaini}, {Edge},
  {Hambly}, {Jameson}, {Lucas}, {Casali}, {Adamson}, {Dye}, {Emerson},
  {Foucaud}, {Hewett}, {Hirst}, {Hodgkin}, {Irwin}, {Lodieu}, {McMahon},
  {Simpson}, {Smail}, {Mortlock}, \& {Folger}}]{Lawrence07}
{Lawrence}, A., {Warren}, S.~J., {Almaini}, O., {et~al.} 2007, \mnras, 379,
  1599

\bibitem[{{Lidz} {et~al.}(2007){Lidz}, {McQuinn}, {Zaldarriaga}, {Hernquist},
  \& {Dutta}}]{Lidz07}
{Lidz}, A., {McQuinn}, M., {Zaldarriaga}, M., {Hernquist}, L., \& {Dutta}, S.
  2007, \apj, 670, 39

\bibitem[{{Luki{\'c}} {et~al.}(2015){Luki{\'c}}, {Stark}, {Nugent}, {White},
  {Meiksin}, \& {Almgren}}]{Lukic15}
{Luki{\'c}}, Z., {Stark}, C.~W., {Nugent}, P., {et~al.} 2015, \mnras, 446, 3697

\bibitem[{{Lusso} {et~al.}(2015){Lusso}, {Worseck}, {Hennawi}, {Prochaska},
  {Vignali}, {Stern}, \& {O'Meara}}]{Lusso15}
{Lusso}, E., {Worseck}, G., {Hennawi}, J.~F., {et~al.} 2015, \mnras, 449, 4204

\bibitem[{{Martini}(2004)}]{Martini04}
{Martini}, P. 2004, Coevolution of Black Holes and Galaxies, 169

\bibitem[{{Mazzucchelli} {et~al.}(2017){Mazzucchelli}, {Ba{\~n}ados},
  {Venemans}, {Decarli}, {Farina}, {Walter}, {Eilers}, {Rix}, {Simcoe},
  {Stern}, {Fan}, {Schlafly}, {De Rosa}, {Hennawi}, {Chambers}, {Greiner},
  {Burgett}, {Draper}, {Kaiser}, {Kudritzki}, {Magnier}, {Metcalfe}, {Waters},
  \& {Wainscoat}}]{Mazzucchelli17}
{Mazzucchelli}, C., {Ba{\~n}ados}, E., {Venemans}, B.~P., {et~al.} 2017, \apj,
  849, 91

\bibitem[{{McGreer} {et~al.}(2015){McGreer}, {Mesinger}, \&
  {D'Odorico}}]{McGreer15}
{McGreer}, I.~D., {Mesinger}, A., \& {D'Odorico}, V. 2015, \mnras, 447, 499

\bibitem[{{McMahon} {et~al.}(2013){McMahon}, {Banerji}, {Gonzalez}, {Koposov},
  {Bejar}, {Lodieu}, {Rebolo}, \& {VHS Collaboration}}]{McMahon13}
{McMahon}, R.~G., {Banerji}, M., {Gonzalez}, E., {et~al.} 2013, The Messenger,
  154, 35

\bibitem[{{Mesinger} \& {Furlanetto}(2007)}]{MF07}
{Mesinger}, A., \& {Furlanetto}, S. 2007, \apj, 669, 663

\bibitem[{{Mesinger} {et~al.}(2011){Mesinger}, {Furlanetto}, \&
  {Cen}}]{Mesinger11}
{Mesinger}, A., {Furlanetto}, S., \& {Cen}, R. 2011, \mnras, 411, 955

\bibitem[{{Mesinger} \& {Furlanetto}(2008)}]{MF08}
{Mesinger}, A., \& {Furlanetto}, S.~R. 2008, \mnras, 385, 1348

\bibitem[{{Mesinger} {et~al.}(2016){Mesinger}, {Greig}, \&
  {Sobacchi}}]{Mesinger16}
{Mesinger}, A., {Greig}, B., \& {Sobacchi}, E. 2016, \mnras, 459, 2342

\bibitem[{{Mesinger} \& {Haiman}(2007)}]{MH07}
{Mesinger}, A., \& {Haiman}, Z. 2007, \apj, 660, 923

\bibitem[{{Miralda-Escud{\'e}}(1998)}]{ME98}
{Miralda-Escud{\'e}}, J. 1998, \apj, 501, 15

\bibitem[{{Mortlock} {et~al.}(2011){Mortlock}, {Warren}, {Venemans}, {Patel},
  {Hewett}, {McMahon}, {Simpson}, {Theuns}, {Gonz{\'a}les-Solares}, {Adamson},
  {Dye}, {Hambly}, {Hirst}, {Irwin}, {Kuiper}, {Lawrence}, \&
  {R{\"o}ttgering}}]{Mortlock11}
{Mortlock}, D.~J., {Warren}, S.~J., {Venemans}, B.~P., {et~al.} 2011, \nat,
  474, 616

\bibitem[{{P{\^a}ris} {et~al.}(2011){P{\^a}ris}, {Petitjean}, {Rollinde},
  {Aubourg}, {Busca}, {Charlassier}, {Delubac}, {Hamilton}, {Le Goff},
  {Palanque-Delabrouille}, {Peirani}, {Pichon}, {Rich}, {Vargas-Maga{\~n}a}, \&
  {Y{\`e}che}}]{Paris11}
{P{\^a}ris}, I., {Petitjean}, P., {Rollinde}, E., {et~al.} 2011, \aap, 530, A50

\bibitem[{{P{\^a}ris} {et~al.}(2017){P{\^a}ris}, {Petitjean}, {Ross}, {Myers},
  {Aubourg}, {Streblyanska}, {Bailey}, {Armengaud}, {Palanque-Delabrouille},
  {Y{\`e}che}, {Hamann}, {Strauss}, {Albareti}, {Bovy}, {Bizyaev}, {Niel
  Brandt}, {Brusa}, {Buchner}, {Comparat}, {Croft}, {Dwelly}, {Fan},
  {Font-Ribera}, {Ge}, {Georgakakis}, {Hall}, {Jiang}, {Kinemuchi},
  {Malanushenko}, {Malanushenko}, {McMahon}, {Menzel}, {Merloni}, {Nandra},
  {Noterdaeme}, {Oravetz}, {Pan}, {Pieri}, {Prada}, {Salvato}, {Schlegel},
  {Schneider}, {Simmons}, {Viel}, {Weinberg}, \& {Zhu}}]{Paris17}
{P{\^a}ris}, I., {Petitjean}, P., {Ross}, N.~P., {et~al.} 2017, \aap, 597, A79

\bibitem[{Pedregosa {et~al.}(2011)Pedregosa, Varoquaux, Gramfort, Michel,
  Thirion, Grisel, Blondel, Prettenhofer, Weiss, Dubourg, Vanderplas, Passos,
  Cournapeau, Brucher, Perrot, \& Duchesnay}]{scikit-learn}
Pedregosa, F., Varoquaux, G., Gramfort, A., {et~al.} 2011, Journal of Machine
  Learning Research, 12, 2825

\bibitem[{{Planck Collaboration} {et~al.}(2016{\natexlab{a}}){Planck
  Collaboration}, {Aghanim}, {Ashdown}, {Aumont}, {Baccigalupi}, {Ballardini},
  {Banday}, {Barreiro}, {Bartolo}, {Basak}, {Battye}, {Benabed}, {Bernard},
  {Bersanelli}, {Bielewicz}, {Bock}, {Bonaldi}, {Bonavera}, {Bond}, {Borrill},
  {Bouchet}, {Boulanger}, {Bucher}, {Burigana}, {Butler}, {Calabrese},
  {Cardoso}, {Carron}, {Challinor}, {Chiang}, {Colombo}, {Combet}, {Comis},
  {Coulais}, {Crill}, {Curto}, {Cuttaia}, {Davis}, {de Bernardis}, {de Rosa},
  {de Zotti}, {Delabrouille}, {Delouis}, {Di Valentino}, {Dickinson}, {Diego},
  {Dor{\'e}}, {Douspis}, {Ducout}, {Dupac}, {Efstathiou}, {Elsner},
  {En{\ss}lin}, {Eriksen}, {Falgarone}, {Fantaye}, {Finelli}, {Forastieri},
  {Frailis}, {Fraisse}, {Franceschi}, {Frolov}, {Galeotta}, {Galli}, {Ganga},
  {G{\'e}nova-Santos}, {Gerbino}, {Ghosh}, {Gonz{\'a}lez-Nuevo}, {G{\'o}rski},
  {Gratton}, {Gruppuso}, {Gudmundsson}, {Hansen}, {Helou},
  {Henrot-Versill{\'e}}, {Herranz}, {Hivon}, {Huang}, {Ili{\'c}}, {Jaffe},
  {Jones}, {Keih{\"a}nen}, {Keskitalo}, {Kisner}, {Knox}, {Krachmalnicoff},
  {Kunz}, {Kurki-Suonio}, {Lagache}, {Lamarre}, {Langer}, {Lasenby},
  {Lattanzi}, {Lawrence}, {Le Jeune}, {Leahy}, {Levrier}, {Liguori}, {Lilje},
  {L{\'o}pez-Caniego}, {Ma}, {Mac{\'{\i}}as-P{\'e}rez}, {Maggio}, {Mangilli},
  {Maris}, {Martin}, {Mart{\'{\i}}nez-Gonz{\'a}lez}, {Matarrese}, {Mauri},
  {McEwen}, {Meinhold}, {Melchiorri}, {Mennella}, {Migliaccio},
  {Miville-Desch{\^e}nes}, {Molinari}, {Moneti}, {Montier}, {Morgante}, {Moss},
  {Mottet}, {Naselsky}, {Natoli}, {Oxborrow}, {Pagano}, {Paoletti},
  {Partridge}, {Patanchon}, {Patrizii}, {Perdereau}, {Perotto}, {Pettorino},
  {Piacentini}, {Plaszczynski}, {Polastri}, {Polenta}, {Puget}, {Rachen},
  {Racine}, {Reinecke}, {Remazeilles}, {Renzi}, {Rocha}, {Rossetti}, {Roudier},
  {Rubi{\~n}o-Mart{\'{\i}}n}, {Ruiz-Granados}, {Salvati}, {Sandri},
  {Savelainen}, {Scott}, {Sirri}, {Sunyaev}, {Suur-Uski}, {Tauber}, {Tenti},
  {Toffolatti}, {Tomasi}, {Tristram}, {Trombetti}, {Valiviita}, {Van Tent},
  {Vibert}, {Vielva}, {Villa}, {Vittorio}, {Wandelt}, {Watson}, {Wehus},
  {White}, {Zacchei}, \& {Zonca}}]{Planck16a}
{Planck Collaboration}, {Aghanim}, N., {Ashdown}, M., {et~al.}
  2016{\natexlab{a}}, \aap, 596, A107

\bibitem[{{Planck Collaboration} {et~al.}(2016{\natexlab{b}}){Planck
  Collaboration}, {Adam}, {Aghanim}, {Ashdown}, {Aumont}, {Baccigalupi},
  {Ballardini}, {Banday}, {Barreiro}, {Bartolo}, {Basak}, {Battye}, {Benabed},
  {Bernard}, {Bersanelli}, {Bielewicz}, {Bock}, {Bonaldi}, {Bonavera}, {Bond},
  {Borrill}, {Bouchet}, {Boulanger}, {Bucher}, {Burigana}, {Calabrese},
  {Cardoso}, {Carron}, {Chiang}, {Colombo}, {Combet}, {Comis}, {Couchot},
  {Coulais}, {Crill}, {Curto}, {Cuttaia}, {Davis}, {de Bernardis}, {de Rosa},
  {de Zotti}, {Delabrouille}, {Di Valentino}, {Dickinson}, {Diego}, {Dor{\'e}},
  {Douspis}, {Ducout}, {Dupac}, {Elsner}, {En{\ss}lin}, {Eriksen}, {Falgarone},
  {Fantaye}, {Finelli}, {Forastieri}, {Frailis}, {Fraisse}, {Franceschi},
  {Frolov}, {Galeotta}, {Galli}, {Ganga}, {G{\'e}nova-Santos}, {Gerbino},
  {Ghosh}, {Gonz{\'a}lez-Nuevo}, {G{\'o}rski}, {Gruppuso}, {Gudmundsson},
  {Hansen}, {Helou}, {Henrot-Versill{\'e}}, {Herranz}, {Hivon}, {Huang},
  {Ili{\'c}}, {Jaffe}, {Jones}, {Keih{\"a}nen}, {Keskitalo}, {Kisner}, {Knox},
  {Krachmalnicoff}, {Kunz}, {Kurki-Suonio}, {Lagache}, {L{\"a}hteenm{\"a}ki},
  {Lamarre}, {Langer}, {Lasenby}, {Lattanzi}, {Lawrence}, {Le Jeune},
  {Levrier}, {Lewis}, {Liguori}, {Lilje}, {L{\'o}pez-Caniego}, {Ma},
  {Mac{\'{\i}}as-P{\'e}rez}, {Maggio}, {Mangilli}, {Maris}, {Martin},
  {Mart{\'{\i}}nez-Gonz{\'a}lez}, {Matarrese}, {Mauri}, {McEwen}, {Meinhold},
  {Melchiorri}, {Mennella}, {Migliaccio}, {Miville-Desch{\^e}nes}, {Molinari},
  {Moneti}, {Montier}, {Morgante}, {Moss}, {Naselsky}, {Natoli}, {Oxborrow},
  {Pagano}, {Paoletti}, {Partridge}, {Patanchon}, {Patrizii}, {Perdereau},
  {Perotto}, {Pettorino}, {Piacentini}, {Plaszczynski}, {Polastri}, {Polenta},
  {Puget}, {Rachen}, {Racine}, {Reinecke}, {Remazeilles}, {Renzi}, {Rocha},
  {Rossetti}, {Roudier}, {Rubi{\~n}o-Mart{\'{\i}}n}, {Ruiz-Granados},
  {Salvati}, {Sandri}, {Savelainen}, {Scott}, {Sirri}, {Sunyaev}, {Suur-Uski},
  {Tauber}, {Tenti}, {Toffolatti}, {Tomasi}, {Tristram}, {Trombetti},
  {Valiviita}, {Van Tent}, {Vielva}, {Villa}, {Vittorio}, {Wandelt}, {Wehus},
  {White}, {Zacchei}, \& {Zonca}}]{Planck16b}
{Planck Collaboration}, {Adam}, R., {Aghanim}, N., {et~al.} 2016{\natexlab{b}},
  \aap, 596, A108

\bibitem[{{Rahmati} {et~al.}(2013){Rahmati}, {Pawlik}, {Rai{\v c}evi{\'c}}, \&
  {Schaye}}]{Rahmati13}
{Rahmati}, A., {Pawlik}, A.~H., {Rai{\v c}evi{\'c}}, M., \& {Schaye}, J. 2013,
  \mnras, 430, 2427

\bibitem[{{Reed} {et~al.}(2017){Reed}, {McMahon}, {Martini}, {Banerji},
  {Auger}, {Hewett}, {Koposov}, {Gibbons}, {Gonzalez-Solares}, {Ostrovski},
  {Tie}, {Abdalla}, {Allam}, {Benoit-L{\'e}vy}, {Bertin}, {Brooks},
  {Buckley-Geer}, {Burke}, {Carnero Rosell}, {Carrasco Kind}, {Carretero}, {da
  Costa}, {DePoy}, {Desai}, {Diehl}, {Doel}, {Evrard}, {Finley}, {Flaugher},
  {Fosalba}, {Frieman}, {Garc{\'{\i}}a-Bellido}, {Gaztanaga}, {Goldstein},
  {Gruen}, {Gruendl}, {Gutierrez}, {James}, {Kuehn}, {Kuropatkin}, {Lahav},
  {Lima}, {Maia}, {Marshall}, {Melchior}, {Miller}, {Miquel}, {Nord}, {Ogando},
  {Plazas}, {Romer}, {Sanchez}, {Scarpine}, {Schubnell}, {Sevilla-Noarbe},
  {Smith}, {Sobreira}, {Suchyta}, {Swanson}, {Tarle}, {Tucker}, {Walker}, \&
  {Wester}}]{Reed17}
{Reed}, S.~L., {McMahon}, R.~G., {Martini}, P., {et~al.} 2017, \mnras, 468,
  4702

\bibitem[{{Robertson} {et~al.}(2015){Robertson}, {Ellis}, {Furlanetto}, \&
  {Dunlop}}]{Robertson15}
{Robertson}, B.~E., {Ellis}, R.~S., {Furlanetto}, S.~R., \& {Dunlop}, J.~S.
  2015, \apjl, 802, L19

\bibitem[{{Rorai} {et~al.}(2013){Rorai}, {Hennawi}, \& {White}}]{Rorai13}
{Rorai}, A., {Hennawi}, J.~F., \& {White}, M. 2013, \apj, 775, 81

\bibitem[{{Schmidt} {et~al.}(2017{\natexlab{a}}){Schmidt}, {Hennawi},
  {Worseck}, {Davies}, {Luki{\'c}}, \& {O{\~n}orbe}}]{Schmidt17b}
{Schmidt}, T.~M., {Hennawi}, J.~F., {Worseck}, G., {et~al.} 2017{\natexlab{a}},
  ArXiv e-prints, arXiv:1710.04527

\bibitem[{{Schmidt} {et~al.}(2017{\natexlab{b}}){Schmidt}, {Worseck},
  {Hennawi}, {Prochaska}, \& {Crighton}}]{Schmidt17a}
{Schmidt}, T.~M., {Worseck}, G., {Hennawi}, J.~F., {Prochaska}, J.~X., \&
  {Crighton}, N.~H.~M. 2017{\natexlab{b}}, \apj, 847, 81

\bibitem[{{Schroeder} {et~al.}(2013){Schroeder}, {Mesinger}, \&
  {Haiman}}]{Schroeder13}
{Schroeder}, J., {Mesinger}, A., \& {Haiman}, Z. 2013, \mnras, 428, 3058

\bibitem[{{Simcoe} {et~al.}(2012){Simcoe}, {Sullivan}, {Cooksey}, {Kao},
  {Matejek}, \& {Burgasser}}]{Simcoe12}
{Simcoe}, R.~A., {Sullivan}, P.~W., {Cooksey}, K.~L., {et~al.} 2012, \nat, 492,
  79

\bibitem[{{Sorini} {et~al.}(2017){Sorini}, {O{\~n}orbe}, {Hennawi}, \&
  {Luki{\'c}}}]{Sorini17}
{Sorini}, D., {O{\~n}orbe}, J., {Hennawi}, J.~F., \& {Luki{\'c}}, Z. 2017,
  ArXiv e-prints, arXiv:1709.03988

\bibitem[{{Stevans} {et~al.}(2014){Stevans}, {Shull}, {Danforth}, \&
  {Tilton}}]{Stevans14}
{Stevans}, M.~L., {Shull}, J.~M., {Danforth}, C.~W., \& {Tilton}, E.~M. 2014,
  \apj, 794, 75

\bibitem[{{Suzuki} {et~al.}(2005){Suzuki}, {Tytler}, {Kirkman}, {O'Meara}, \&
  {Lubin}}]{Suzuki05}
{Suzuki}, N., {Tytler}, D., {Kirkman}, D., {O'Meara}, J.~M., \& {Lubin}, D.
  2005, \apj, 618, 592

\bibitem[{{Telfer} {et~al.}(2002){Telfer}, {Zheng}, {Kriss}, \&
  {Davidsen}}]{Telfer02}
{Telfer}, R.~C., {Zheng}, W., {Kriss}, G.~A., \& {Davidsen}, A.~F. 2002, \apj,
  565, 773

\bibitem[{{Venemans} {et~al.}(2013){Venemans}, {Findlay}, {Sutherland}, {De
  Rosa}, {McMahon}, {Simcoe}, {Gonz{\'a}lez-Solares}, {Kuijken}, \&
  {Lewis}}]{Venemans13}
{Venemans}, B.~P., {Findlay}, J.~R., {Sutherland}, W.~J., {et~al.} 2013, \apj,
  779, 24

\bibitem[{{Venemans} {et~al.}(2015){Venemans}, {Ba{\~n}ados}, {Decarli},
  {Farina}, {Walter}, {Chambers}, {Fan}, {Rix}, {Schlafly}, {McMahon},
  {Simcoe}, {Stern}, {Burgett}, {Draper}, {Flewelling}, {Hodapp}, {Kaiser},
  {Magnier}, {Metcalfe}, {Morgan}, {Price}, {Tonry}, {Waters}, {AlSayyad},
  {Banerji}, {Chen}, {Gonz{\'a}lez-Solares}, {Greiner}, {Mazzucchelli},
  {McGreer}, {Miller}, {Reed}, \& {Sullivan}}]{Venemans15}
{Venemans}, B.~P., {Ba{\~n}ados}, E., {Decarli}, R., {et~al.} 2015, \apjl, 801,
  L11

\bibitem[{{Venemans} {et~al.}(2017{\natexlab{a}}){Venemans}, {Walter},
  {Decarli}, {Ba{\~n}ados}, {Carilli}, {Winters}, {Schuster}, {da Cunha},
  {Fan}, {Farina}, {Mazzucchelli}, {Rix}, \& {Weiss}}]{Venemans17b}
{Venemans}, B.~P., {Walter}, F., {Decarli}, R., {et~al.} 2017{\natexlab{a}},
  \apjl, 851, L8

\bibitem[{{Venemans} {et~al.}(2017{\natexlab{b}}){Venemans}, {Walter},
  {Decarli}, {Ba{\~n}ados}, {Hodge}, {Hewett}, {McMahon}, {Mortlock}, \&
  {Simpson}}]{Venemans17}
---. 2017{\natexlab{b}}, \apj, 837, 146

\bibitem[{{Wang} {et~al.}(2017){Wang}, {Fan}, {Yang}, {Wu}, {Yang}, {Bian},
  {McGreer}, {Li}, {Li}, {Ding}, {Dey}, {Dye}, {Findlay}, {Green}, {James},
  {Jiang}, {Lang}, {Lawrence}, {Myers}, {Ross}, {Schlegel}, \&
  {Shanks}}]{Wang17}
{Wang}, F., {Fan}, X., {Yang}, J., {et~al.} 2017, \apj, 839, 27

\bibitem[{{White} {et~al.}(2003){White}, {Becker}, {Fan}, \&
  {Strauss}}]{White03}
{White}, R.~L., {Becker}, R.~H., {Fan}, X., \& {Strauss}, M.~A. 2003, \aj, 126,
  1

\bibitem[{{Wyithe} {et~al.}(2005){Wyithe}, {Loeb}, \& {Carilli}}]{Wyithe05}
{Wyithe}, J.~S.~B., {Loeb}, A., \& {Carilli}, C. 2005, \apj, 628, 575

\bibitem[{{Young} {et~al.}(1979){Young}, {Sargent}, {Boksenberg}, {Carswell},
  \& {Whelan}}]{Young79}
{Young}, P.~J., {Sargent}, W.~L.~W., {Boksenberg}, A., {Carswell}, R.~F., \&
  {Whelan}, J.~A.~J. 1979, \apj, 229, 891

\bibitem[{{Zel'dovich}(1970)}]{Zel'dovich70}
{Zel'dovich}, Y.~B. 1970, \aap, 5, 84

\end{thebibliography}
\end{document}